\documentclass[prb,aps,twocolumn,groupedaddress,floats,showpacs,final,superscriptaddress]{revtex4-1}
\usepackage[latin3]{inputenc}
\usepackage[makeroom]{cancel}
\usepackage{graphicx}
\usepackage{amsmath}
\usepackage{amsfonts}
\usepackage{amssymb}
\usepackage{color}
\usepackage[left=2cm,right=2cm,top=2cm,bottom=2cm]{geometry}

\usepackage{graphicx} 
\usepackage{dcolumn}
\usepackage{bm}
\usepackage{simplewick}
\usepackage{array}
\usepackage{appendix}

\RequirePackage[
   hyperindex,colorlinks,bookmarksnumbered,
   plainpages=true,pdfstartview=FitH]{hyperref}
\hypersetup{linkcolor=blue,urlcolor=blue,citecolor=blue}
\usepackage{hyperref}

\definecolor{purple}{rgb}{0.5,0,0.6}

\usepackage{ulem}
\renewcommand{\emph}[1]{\textit{#1}}
\definecolor{darkblue}{rgb}{0,0,0.5}
\definecolor{darkgreen}{rgb}{0,0.5,0}
\definecolor{darkred}{rgb}{.7,0,0}
\definecolor{purple}{rgb}{0.5,0,0.6}
\definecolor{orange}{rgb}{1,0.5,0}
\definecolor{grey}{rgb}{.6,.6,.6}
\definecolor{lightpink}{rgb}{1,0.7,0.75}
\definecolor{pink}{rgb}{1,0.4,0.58}
\definecolor{deeppink}{rgb}{1,0.08,0.58}

\renewcommand{\emph}[1]{\textit{#1}}

\begin{document}


\title{{\color{black} Thermoelectric Transport through SU(N) Kondo Impurity}}



\author{D. B. Karki}
\affiliation{The  Abdus  Salam  International  Centre  for  Theoretical  Physics  (ICTP),
Strada  Costiera 11, I-34151  Trieste,  Italy}
\affiliation{School for Advanced Studies (SISSA), Via Bonomea 265, 34136 Trieste, Italy}
\author{Mikhail N. Kiselev}
\affiliation{The  Abdus  Salam  International  Centre  for  Theoretical  Physics  (ICTP),
Strada  Costiera 11, I-34151  Trieste,  Italy}



\begin{abstract}
We investigate thermoelectric transport through a SU(N) quantum impurity in the Kondo regime. The strong coupling fixed point theory is described by the local Fermi-liquid paradigm. Using Keldysh technique we analyse the electric current through the quantum impurity at both finite bias voltage and finite temperature drop across it.  The theory of a steady state at zero-current provides a complete description of the Seebeck effect.  We find pronounced non-linear effects in temperature drop at low temperatures.
We illustrate the significance of the non-linearities for enhancement of thermopower by two examples of SU(4) symmetric regimes characterized by a filling factor $m$: i) particle-hole symmetric at 
$m$$=$$2$ and ii) particle-hole non-symmetric at $m$$=$$1$. We analyse the effects of potential scattering and coupling asymmetry on the transport coefficients. We discuss connections between the theory and transport experiments with coupled quantum dots and carbon nanotubes.
\end{abstract}

\maketitle

\noindent 
{\it Introduction.}  Recent progress in understanding of thermoelectric phenomena on the nanoscale
stimulated both new experiments \cite{Pierre_NAT(536)_2016, Pierre_NAT(526)_2015,Pierre_SCI(342)_2013} and development of 
new theoretical approaches to this problem (see e.g. \cite{Thermo_review} for review). One of the fundamental properties of the quantum transport through nano-sized objects (quantum dots (QD), carbon nanotubes (CNT), quantum point contacts (QPC) etc) is associated with the charge quantization \cite{Blanter_Nazarov}.
It offers a very efficient tool for the quantum manipulation of the single-electron devices being  building blocks for  quantum information processing. The universality of the heat flows in the quantum regime, scales of the quantum interference effects and limits of the tunability are the central questions of the new emergent field of the quantum heat transport \cite{Pierre_NAT(536)_2016, Pierre_NAT(526)_2015,Pierre_SCI(342)_2013, Molenkamp_1990, Molenkamp_1992, Scheibner_PRL(95)_2005}. Besides, the effects of strong electron correlations and resonance scattering become very pronounced at low temperatures and can be measured with high controllability (e.g. external electric and magnetic fields, geometry, temperature etc) of the semiconductor nano-devices. Therefore, investigation of the quantum effects and influence of strong correlations and resonance scattering on the heat transport (both experimentally and theoretically) is one of the cornerstones of quantum electronics. 

\begin{figure}[b]
\begin{center}
 \includegraphics[width=75mm]{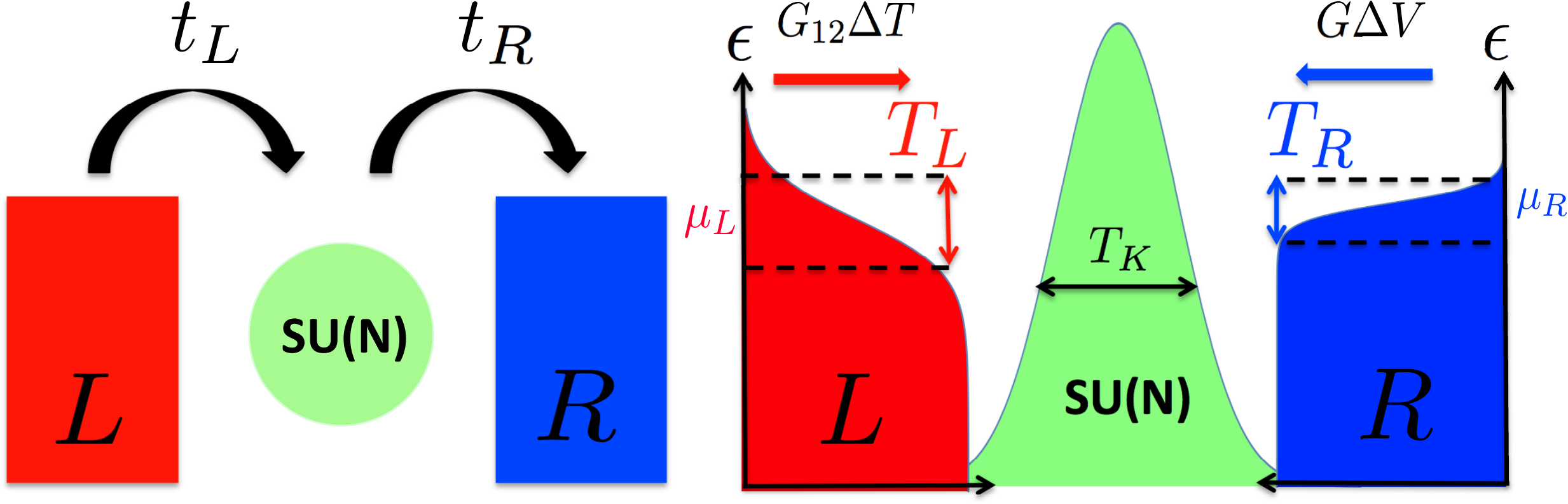} 
\caption{(Color online) Left panel: cartoon for the tunneling $t_{L/R}$ through the SU(N) quantum impurity (see the main text). Right panel: Fermi distribution functions of the left (hot) and right (cold) leads at temperatures $T_{L/R}$ and chemical potential $\mu_{L/R}$. The thermo-voltage $\Delta V=|\mu_R-\mu_L|/e$ is applied to achieve the steady state with zero net current across the impurity. Left (red)/right (blue) arrows show directions of thermo- and electric- currents. Resonance Kondo peak of width $T_K$ in DoS is shown by the green color.}
\label{f.1}
\end{center} 
\end{figure}

As  follows from the Fermi-liquid (FL) theory, the thermoelectric power (Seebeck effect) of bulk metals is directly proportional to the temperature $T$ and inversely proportional to the Fermi energy $\varepsilon_F$ \cite{Zlatic}. The resonance scattering on a quantum impurity, however, dramatically enhances this effect due to the emergence of new quasiparticle resonances at the Fermi level described by the Kondo effect 
\cite{Nozieres, Nozieres_Blandin_JPhys_1980, Hewson}. The contribution to the Seebeck effect proportional to the concentration of impurities at low $T$, as a result, scales as $T/T_K$ \cite{Hewson, Zlatic} where $T_K$ is a characteristic energy defining the width of the Kondo resonance, the Kondo temperature (Fig. \ref{f.1}). The Kondo effect in nano-devices is key for enhancing the thermoelectric transport coefficients \cite{Scheibner_PRL(95)_2005}. The tunable thermo-transport through nano-devices controlling the heat flow is needed for efficient operation of quantum circuits elements: single-electron transistors, quantum diodes etc to perform controllable heat guiding.

The Kondo effect has been observed in the experiments on the semiconductor quantum dots and the single wall carbon nanotubes \cite{Goldhaber_nat(391)_1998,Cronewett_SCI(281)_1998,Jesper_NAT(408)_2000, Ludwig_2006}. 
The effect manifests itself
by complete screening of spin of the quantum impurity and, as a result, the FL behaviour in the strong coupling (low temperatures) regime \cite{Nozieres, Nozieres_Blandin_JPhys_1980, Affleck_Lud_PRB(48)_1993}.
Here we use the local FL paradigm 
\cite{Mora_PRB_(80)_2009,Vitushinsky_Clerk_Hur_PRL_(100)_2008,Mora_Leyronas_Regnault_PRL_(100)_2008,
Mora_Clerk_Hur_PRB(80)_2009,Mora_Moca_Delft_Zarand_PRB(92)_2015,Hur_Simon_Loss_PRB(75)_2007, Mora_Schuricht_(89)_2014, HWDK_PRB_(89)_2014} which is a powerful tool for the description of thermodynamic and transport properties of quantum impurity in the strong coupling regime. It has also been applied recently for explanation of
"0.7-anomaly" in QPCs \cite{Meir_NAT_(442)_2006,JVD_NAT_(501)_2013}.
The $s$$=$$1/2$ SU(2) Kondo impurity physics arises at the half-filled particle-hole (PH) symmetric regime.
We refer to "electrons" as quasiparticles above $\varepsilon_F$ and "holes" as the excitations below 
$\varepsilon_F$. The PH symmetry, being responsible for the enhancement of the electric conductance, suppresses however the thermo-electric transport: the {\color{black} thermo-} current carried by electrons is completely compensated by heat current carried by holes challenging however to investigate Kondo models in the regime away from PH-symmetry.
To achieve appreciable thermopower, the occupation factor of the quantum impurity should be integer
(Coulomb blockade valleys \cite{Blanter_Nazarov}), while the particle-hole symmetry should be lifted. Such properties are generic for the SU(N) Kondo models with the filling factors different from $1/2$.  

The SU(N) Kondo physics with $N$$=$$4$ is experimentally realized in CNTs
\cite{Jarillo-Herrero_Nat(434)_2005, Finkelstein_PRB(75)_2007, Makarovski_Liu_Finkelstein_PRL(99)_2007,
Ferrier_NAT_(12)_2016} and {\color{black} double} QDs \cite{Keller_Goldhaber_nat_phys(10)_2014}. The SU(N) Kondo model has also been {\color{black} proposed to investigate} in ultra-cold gases experiments \cite{Bauer_Salomon_Demler_PRL_(111)_2013,Nishida_PRL_(111)_2013}. There are several theoretical suggestions 
for realization of SU(N) Kondo physics with {\color{black} $N$$=$$3$ \cite{Nishida_PRL_(111)_2013,Nishida_PRA_(93)_2016}}
$N$$=$$6$ \cite{Kuzmenko_Avishai_2016} and $N$$=$$12$ reported in \cite{Avishai_PRB_(89)_2014}. While the electron transport through SU(N) Kondo impurity is well understood theoretically
\cite{Hur_Simon_Borda_PRB_(69)_2004,Lopez_Hur_PRB_(71)_2005,Schmid_Grifoni_PRB_(91)_2015,
Mora_Clerk_Hur_PRB(80)_2009}, the thermo-electric transport in the Kondo regime {\color{black} remains challenging} \cite{Sakano_JMMM_(310)_2006,Sakano_JPSJ_(76)_2007,Tosi_PRB_(86)_2012, Azema_PRB(86)_2012,
Andergassen_PRB_(94)_2016}. 

In this {\it Letter} we present
a full fledged theory for the Seebeck effect of SU(N) Kondo model for the strong coupling regime 
$T$$\ll $$T_K$. Our approach is based on real time out-of equilibrium Keldysh calculations. 
We used the local Fermi-liquid paradigm for constructing 
a perturbative expansion for the electric current around the strong coupling fixed point of the model. 
We illustrate the thermoelectric properties of the SU(N) Kondo model on two particular examples, namely,
$N$$=$$4$ with the filling factors $1/4$ and $1/2$. We compute the thermoelectric power for arbitrary
temperature drop between the electron reservoirs and discuss the significance of non-linear effects in temperature drop.

{\it Setup.} We consider an SU(N) quantum impurity (such as a CNT or coupled QDs) 
sandwiched between
two leads (Fig. \ref{f.1}). The model geometry resembles the experimental setup \cite{Scheibner_PRL(95)_2005}. The temperature of the drain electrode (R) is taken as the reference temperature of the system. The temperature of the source electrode (L) is controlled by the Joule heat released due to the finite current flowing along the lead \cite{Scheibner_PRL(95)_2005}.
Thus, the temperature drop $\Delta T$ is fixed for all measurements. The bias voltage $\Delta V$ is applied between the source and the drain in order to stop the thermo-current (Fig. \ref{f.1} right panel):
\begin{eqnarray}
I=0=G(T)\Delta V+G_{12}(T)\Delta T.
\label{current_exp}
\end{eqnarray}
The differential thermoelectric power is defined at the total current across the impurity tuned to zero:
\begin{eqnarray}
S(T)=\left. -\lim_{\Delta T\to 0}\frac{\Delta V}{\Delta T} \right|_{I=0}=\frac{G_{12}(T)}{G(T)},
\label{thermo_eexp}
\end{eqnarray}
$G$$=$$\partial{I}$$/$$\partial{\Delta V}$$\left. \right|_{\Delta T=0}$ is the electric conductance and  $G_{12}$$=$$\partial{I}$$/$$\partial{\Delta T}$$\left. \right|_{\Delta V=0}$  is the thermoelectric coefficient.

{\it Model.} The tunneling of electrons through the SU(N) quantum impurity (Fig.\ref{f.1} left panel) is described by the Anderson model \cite{Mora_Clerk_Hur_PRB(80)_2009}:
\begin{equation}\label{HamAnd}
\begin{split}
H=&\sum_{k\alpha r}\varepsilon_k c^\dagger_{\alpha k r}c_{\alpha k r}+\sum_{r} \epsilon_{0} d^\dagger_{r}d_{r}\\
&+\sum_{r < r'}Ud^\dagger_{r}d_{r}d^\dagger_{r'}d_{r'}+\sum_{k\alpha r}t_{\alpha}d^\dagger_{r}c_{\alpha k r}+H.c.
\end{split}
\end{equation}
\noindent Here $d_{r}$ annihilates an electron in the dot level $\epsilon_{0}$ with orbitals $r=1, 2, ...N$, $c_{\alpha k r}$ annihilates a conduction electron with the momentum $k$ and orbital $r$ in the leads 
$\alpha=L, R$ and $U$ is the Coulomb repulsion (charging) energy in the dot, $t_\alpha$ is lead-dot tunneling and
$\varepsilon_k=\epsilon_k-\varepsilon_F$ is the linearized conductance electron's dispersion. We assume that the charging energy $U$ is the largest energy scale of the model and therefore take into account only "last" occupied state.
We project out the charge states by  
applying the Schrieffer-Wolff transformation \cite{Schrieffer_Wolf_PR(149)_1966}. As a result we obtain the effective  SU(N) Kondo model describing the physics at the weak coupling $T$$\gg $$T_K$ limit:
\noindent
\begin{eqnarray}\label{HamKon}
\mathcal{H}_K=J_K^{\alpha\beta} \left(\bf{c^\dagger_\alpha \lambda^{\mu}c_{\beta}}\right)\left(\bf{d^\dagger \Lambda^{\mu}d}\right),
\end{eqnarray}
\noindent where $\bf{c^\dagger}$=$(c^\dagger_1,....,c^\dagger_N)$ is a row vector of the electron states in the leads and $\bf{d^\dagger}$=$(d^\dagger_1,....,d^\dagger_N)$ represents the local states in the dot. The $SU(N)$ generators $\bf{\lambda^{\mu}}$ and $\bf{\Lambda^{\mu}}$ for $\mu=1, 2,....N^2-1$ are traceless $N\times N$ Hermitian matrices of the fundamental representation, satisfying the commutation relations
$\left[\bf{\lambda^{\mu}}, \bf{\lambda^{\nu}}\right]$$=$$i f^{\mu\nu\rho}\bf{\lambda^{\rho}}$
where $f^{\mu\nu\rho}$ is the set of fully anti-symmetric structure factors.
As a last step we diagonalize the matrix $J_K^{\alpha\beta}$$\sim$$|$$t_\alpha$$t_\beta$$|$$/U$ 
in the sub-space of two leads $\alpha$,$\beta$$=$$L$,$R$ performing the Glazman-Raikh rotation \cite{Glazman_Raikh_JETP_(27)_1988,Ng_Lee_PRL_(61)_1988,GP_Review_2005}.
Similarly to the SU(2) Kondo model, the anti-symmetric combination of the electron states in the leads $a^\dagger=(c^\dagger_L-c^\dagger_R)/\sqrt{2}$ is fully decoupled
from the Hamiltonian while the symmetric combinations  
$b^\dagger=(c^\dagger_L+c^\dagger_R)/\sqrt{2}$ remains coupled to the quantum impurity \footnote{The weak coupling Hamiltonian (\ref{HamKon}) sets the Kondo temperature
$T_K \sim D\exp\left[-1/(N\nu J)\right]$.  Here $D$ is a bandwidth of conduction electrons band.} 
(without loss of generality we present here the results for symmetric $t_L$$=$$t_R$ dot-lead coupling; general equations for arbitrary coupling  are presented in Supplemental Materials \cite{Karki_MK_Supplement}).

The FL Hamiltonian describing the strong coupling $T$$\ll $$T_K$ regime is obtained by applying the standard point-splitting procedure to {\color{black} 
$:\left(\bf{b^\dagger \lambda^\mu b}\right)\cdot \left(\bf{b^\dagger \lambda^\mu b}\right):$} (see \cite{Affleck_Lud_PRB(48)_1993} for the details), $H_{FL}$$=$$H_0$$+$$H_{\alpha}$$+$$H_{\phi}$ \footnote{We omitted the six-fermion term \cite{ Mora_Clerk_Hur_PRB(80)_2009} in (\ref{HamFL}) since it produces perturbative corrections to the current beyond the accuracy of our theory.}:
\begin{eqnarray}
H_{0}&=&\nu\sum_{r}\int_{\varepsilon}  \varepsilon \left[ a^{\dagger}_{\varepsilon r} a_{\varepsilon r}+b^{\dagger}_{\varepsilon r} b_{\varepsilon r}\right]\\
H_{\alpha}&=&-\sum_{r}\int_{\varepsilon_{1-2}}\left[\frac{\alpha_1}{2\pi}(\varepsilon_1+\varepsilon_2)+\frac{\alpha_2}{4\pi}(\varepsilon_1+\varepsilon_2)^2\right]b^{\dagger}_{\varepsilon_1 r} b_{\varepsilon_2 r}
\nonumber\\
H_{\phi}&=&\sum_{r < r'}\int_{\varepsilon_{1-4}}\left[\frac{\phi_1}{\pi\nu}+\frac{\phi_2}{4\pi\nu}\left(\sum^4_{i=1}\varepsilon_i\right)\right]:b^{\dagger}_{\varepsilon_1 r} b_{\varepsilon_2 r} b^{\dagger}_{\varepsilon_3 r'} b_{\varepsilon_4 r'}:.\nonumber
\label{HamFL}
\end{eqnarray}
The {\it Kondo floating} paradigm \cite{Nozieres, Nozieres_Blandin_JPhys_1980, Affleck_Lud_PRB(48)_1993, Mora_PRB_(80)_2009,Vitushinsky_Clerk_Hur_PRL_(100)_2008,Mora_Leyronas_Regnault_PRL_(100)_2008,
Mora_Clerk_Hur_PRB(80)_2009,Mora_Moca_Delft_Zarand_PRB(92)_2015,Hur_Simon_Loss_PRB(75)_2007, Mora_Schuricht_(89)_2014, HWDK_PRB_(89)_2014} leads to the following FL identities: 
$\alpha_1$$=$$($$N$$-$$1$$)\phi_1$
and $\alpha_2$$=$$($$N$$-$$1$$)\phi_2/4$, $\nu$ is the density of states at $\varepsilon_F$. The connection between $\alpha_1$ and $\alpha_2$ is given by the Bethe ansatz
\cite{Mora_Clerk_Hur_PRB(80)_2009}. We use $\alpha_1$$=$$1/T_K$ as the definition of the Kondo temperature \cite{GP_Review_2005}.

{\it Charge Current.}
The current operator at position $x$ is expressed in terms of first-quantized operators $\psi$
attributed to the linear combinations of the Fermi operators in both leads
\begin{eqnarray}\label{aaa}
\hat{I}(x)=\frac{\hbar  e}{2mi}\sum_r \left[\psi^\dagger_r(x)\partial_x\psi_r(x)-\partial_x \psi^\dagger_r(x) \psi_r(x)\right].
\end{eqnarray}
\noindent For the expansion of  Eq.\eqref{aaa}, we choose the basis of scattering states that includes completely elastic and Hartree terms \cite{Mora_Clerk_Hur_PRB(80)_2009} to get it in compact form:
\begin{eqnarray}
\label{bbb}
\hat{I}=\frac{e}{2\nu h}\sum_{r}  \left[a^{\dagger}_{r}(x)b_{r}(x)-a^{\dagger}_{r}(-x)Sb_{r}(-x)+H. c.\right],\;\;\;
\end{eqnarray}
\normalsize
\noindent {\color{black} where $a_r(x)$$=$$\sum_k$$a_{kr}$$e^{ikx}$, $b_r(x)$$=$$\sum_k$$b_{kr}$$e^{ikx}$, 
$Sb_r(x)$$=$$\sum_k$$S_kb_{kr}$$e^{ikx}$ and the $N$$\times $$N$ $S$-matrix is expressed in terms of a phase shift $\delta$$($$\varepsilon_k$$)$ as $S_k$$=e^{2i\delta(\varepsilon_k)}$.}

{\it Elastic current.}
{\color{black} Calculation of the expectation value of (\ref{bbb}) in the absence of interactions}
is equivalent to use the
Landauer-B\"uttiker formalism \cite{Blanter_Nazarov}:
\begin{eqnarray}\label{elci}
I_{el}=\frac{Ne}{h}\int^{\infty}_{-\infty} d \varepsilon \mathcal{T}(\varepsilon) \Delta f(\varepsilon),
\end{eqnarray}
\noindent where $\Delta f(\varepsilon)$$=$$f_L(\varepsilon)$$-$$f_R(\varepsilon)$, $f_{L/R}$ are Fermi distribution functions of L/R leads; $\mu_L$$-$$\mu_R$$=$$e\Delta$$V$$\ll$ $T_K$ are the chemical potentials, 
$T_R$$=$$T$ is the reference temperature and $T_L$$=$$T$$+$$\Delta T$
(Fig.\ref{f.1} right panel).
The energy dependent transmission  $\mathcal{T}(\varepsilon)$$=$$\sin^2\left[\delta(\varepsilon)\right]$. 
Following Ref. \cite{Mora_Clerk_Hur_PRB(80)_2009}, we Taylor-expand the phase shift {\color{black} for all flavours $r$} in the presence of voltage bias $e \Delta V$ and temperature drop $\Delta T$ as
$\delta_r(\varepsilon)=\delta_0+\alpha_1\varepsilon+\alpha_2 \left(\varepsilon^2- {\cal A}\right)$,
where ${\cal A}=\left[\frac{(e\Delta V)^2}{4}+\frac{(\pi T)^2}{3}+\frac{\pi^2 T \Delta T}{3}\right]$
and $\delta_0$$=$$\pi$$m$$/N$ for the quantum impurity's occupation $m$$=$$1$$,...,$$N$$-$$1$.
The zero energy transmission is given by $\mathcal{T}_0$$=$$\sin^2\delta_0$. Using the above equation for the phase shift $\delta_r$ we expand the current up to the second order in $T$$/$$T_K\ll 1$ to get the elastic contribution. The linear response result is:
\begin{eqnarray}\label{112}
\frac{I_{el}}{Ne/h}&=&\left[{\color{black} \sin^2\delta_0}+\frac{\alpha^2_1}{3}(\pi T)^2\cos2\delta_0\right]e\Delta V
\nonumber\\
&-&\left[\frac{\alpha_1}{3T}(\pi T)^2\sin2\delta_0\right]\Delta T.
\end{eqnarray}
\begin{figure}[t]
\begin{center}
 \includegraphics[width=80mm]{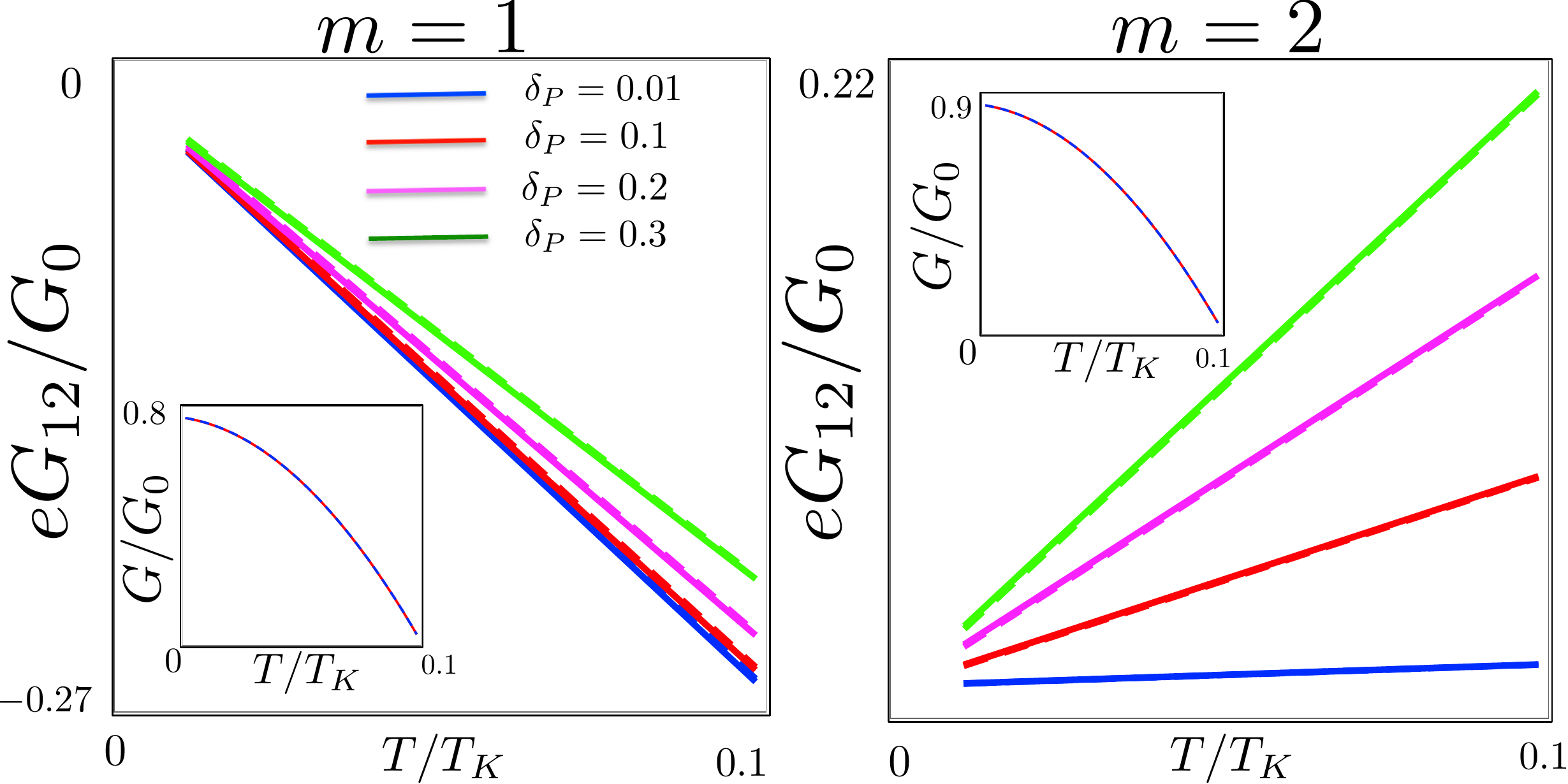}
\caption{(Color online) Main frame: Thermoelectric coefficient $G_{12}$ as a function of the reference temperature $T$$=$$T_R$ at different values of potential scattering $\delta_P$.
Insert: differential conductance $G$ as a function of $T$ for $\delta_P$$=$$0.3$.
The legend is shown on the left panel. Solid lines - numerical solution obtained from (\ref{current_exp}) with (S6-S7, S11-S12)
\cite{Karki_MK_Supplement}. Dashed lines - analytical solution given by (\ref{elecondu}, \ref{thermoc}). Solid and dashed lines are almost indistinguishable.}
\label{f.2}
\end{center} 
\end{figure}
{\it Inelastic current.}
The inelastic contribution to the current is computed using the non-equilibrium Keldysh formalism \cite{Keldysh_SP_JETP(20)_1965}:
\begin{equation}\label{wick}
\delta I_{in}=\langle T_{_C} \hat{I}(t)e^{-i\int dt^{\prime} H_{\phi}(t^{\prime})}\rangle,
\end{equation}
\noindent where $C$ denotes the double side $\eta=\pm$ Keldysh contour \cite{Keldysh_SP_JETP(20)_1965}. Here $T_C$ is the time ordering operator on a contour and the average is performed with the Hamiltonian $H_0$ whereas the contribution from $H_\alpha$ is already accounted in $I_{el}$. 
As discussed in detail in Ref. \cite{Mora_Moca_Delft_Zarand_PRB(92)_2015}, the second order interaction correction to the current is expressed in terms of the self-energies
\begin{eqnarray}\label{lc}
\delta I_{in}=\mathcal{S}\int^{\infty}_{-\infty} \frac{d\varepsilon}{2\pi}\left(\Sigma^{-+}(\varepsilon)-\Sigma^{+-}(\varepsilon)\right) i\pi\nu \Delta f(\varepsilon),\;\;\;\;
\end{eqnarray}
\noindent we used the notation: $\mathcal{S}=\frac{N(N-1)e\pi}{h} \cos2\delta_0$. The self-energies in Eq.\eqref{lc} are defined in terms of the Green's functions as: $\Sigma^{\eta_1, \eta_2}(t)=\left(\frac{\phi_1}{\pi\nu^2}\right)^2\sum_{k_1, k_2, k_3} G^{\eta_1, \eta_2}_{bb}(k_1, t) G^{\eta_2, \eta_1}_{bb}(k_2, -t) G^{\eta_1, \eta_2}_{bb}(k_3, t)$. 
\noindent The {\it local} Green's functions of $aa$$/$$bb$ fermions and the mixed $ab$ fermions in 
{\color{black} real-time} are given by
\begin{eqnarray}\label{dec11}
\mathcal{G}_{\pm}(t)=-\frac{\pi\nu}{2}\left[\frac{T_L e^{-i\mu_L t}}{\sinh(\pi T_L t)}\pm\frac{T_R e^{-i\mu_R t}}{\sinh(\pi T_R t)}\right],
\end{eqnarray}
\noindent with $\mathcal{G}_+(t)$$=$$G^{+-}_{aa/bb}(t)$ and $\mathcal{G}_-(t)$$=$$G_{ba/ab}(t)$$=$$i\pi\nu\Delta f(t)$. Computing the integrals in the presence of the finite bias voltage and finite temperature drop one gets:
\begin{eqnarray}\label{diffself}
&&\Sigma^{-+}(\varepsilon)-\Sigma^{+-}(\varepsilon)=\nonumber\\
&&\frac{\phi^2_1}{i\pi\nu} 
\left[
\frac{3}{4}(e\Delta V)^2+\frac{\Delta T}{T}(\pi T)^2+\varepsilon^2+(\pi T)^2
\right].
\end{eqnarray}
\noindent
In the limit $\Delta$$T$$\to $$0$, $\Delta$$f(\varepsilon)$$ =$$-$$($$\Delta$$ T$$\cdot \varepsilon$$/$$T$$)$$\cdot\partial f/\partial\varepsilon$ and the FL self-energies, being even functions of $\varepsilon$, do  not contribute to the thermo-current {\color{black} at $\Delta V$$=$$0$}. Therefore, the thermoelectric coefficient $G_{12}$ at $\Delta T$$\to $$ 0 $ is fully determined by elastic processes \footnote{Approaches based on the self-energies $\Sigma$ and $T$- matrix are equivalent since 
$\Sigma(\epsilon)=T(\epsilon)(1+{\cal G}_0(\epsilon)\cdot T(\epsilon))^{-1}$, see \cite{Karki_MK_Supplement} for details.}. The linear response inelastic contribution to the current at $\Delta$$T$$=0$ reads
\begin{eqnarray}\label{in112}
\frac{ \delta I_{in}}{Ne/h}  =\left[\frac{2}{3}\left(\pi T\right)^2\left(N-1\right)\phi^2_1 \cos2\delta_0 \right]e\Delta V.
\end{eqnarray}
The equation for the total current beyond the linear response is cumbersome and given in Supplemental Materials \cite{Karki_MK_Supplement}.
Finally, the differential conductance $G$ and differential thermo-electric coefficient
$G_{12}$ are give by:
\begin{eqnarray}\label{elecondu}
G(T)&=&G_0\left[{\color{black} \sin^2\delta_0}+\frac{\alpha^2_1}{3}\frac{N+1}{N-1}(\pi T)^2\cos2\delta_0 \right],\;\;\;\\
\label{thermoc}
G_{12}(T)&=&-G_0\left[\frac{\alpha_1}{3e}\pi^2 T \sin2\delta_0 \right],
\end{eqnarray}
where $G_0=Ne^2/h$ is the unitary conductance.
\begin{figure}[t]
\begin{center}
 \includegraphics[width=80mm]{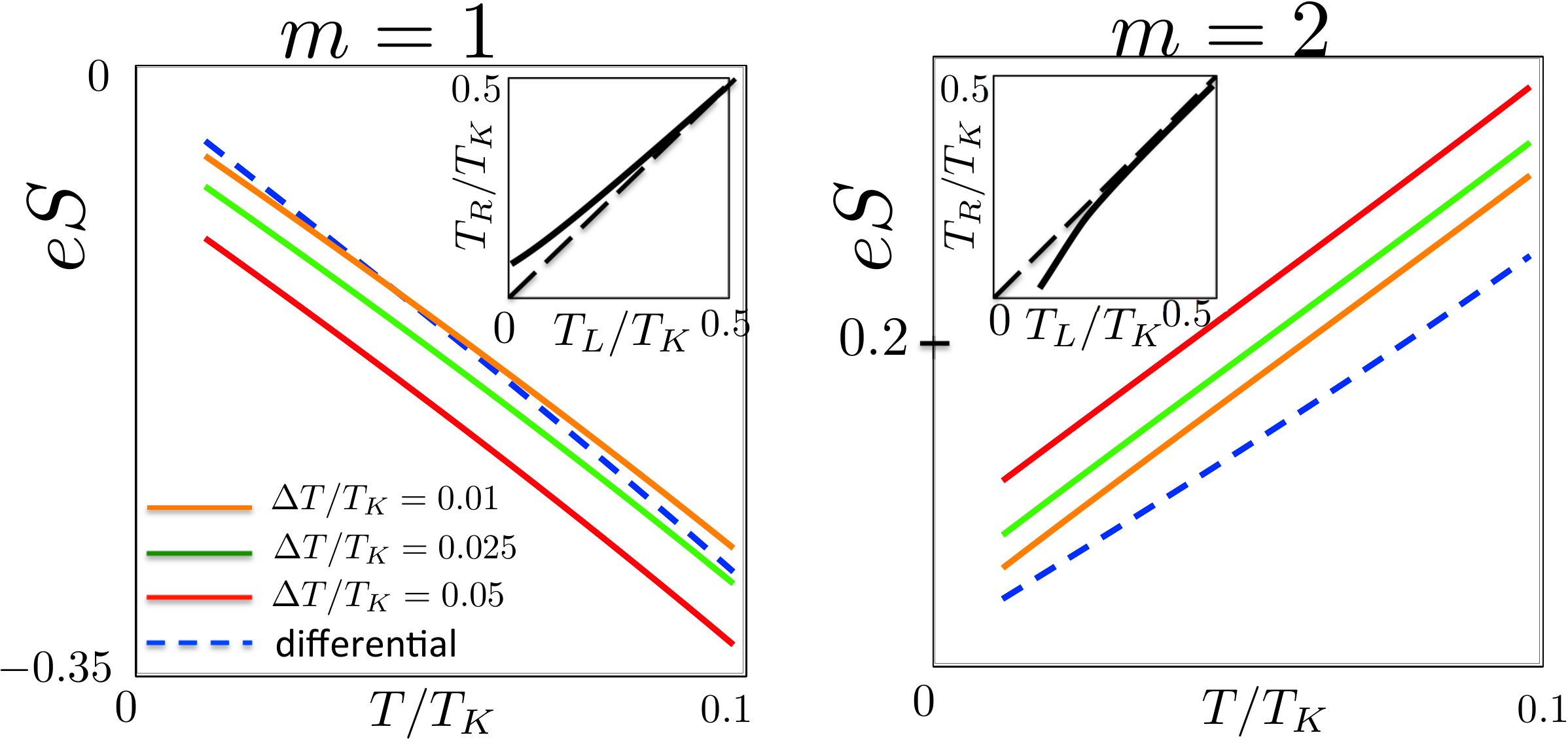}
\caption{(Color online) Main frame: Thermoelectric power $S$ as a function of the reference temperature 
$T$$=$$T_R$; blue dashed curve: differential $S$ given by (\ref{thermo_eexp}) with (\ref{elecondu}, \ref{thermoc}); solid curves correspond to $S$ defined under zero-current condition by (\ref{current_exp}) with (S6-S7,
S11-S12) \cite{Karki_MK_Supplement}
at different values of $\Delta T$ (see the legend); $\delta_P$$=$$0.3$. 
Insert: evolution of the zero current steady state as a function of the temperatures of L-R leads
at finite voltage $e$$\Delta$$V$$/$$T_K$$=0.02$.}
\label{f.3}
\end{center} 
\end{figure}

{\it Potential scattering.} As is seen from (\ref{thermoc}), the $G_{12}$ at given reference temperature $T$ in linear response is proportional to $\sin 2\delta_0$.  For the particle-hole symmetric (PHS) case in the absence of potential scattering $m$$=$$N/2$ (we assume $N$ even), $\delta_0$$=$$\pi/2$  and both $G_{12}$ and the thermoelectric power are zero - the particle thermo current exactly compensates the hole thermo current (Fig. \ref{f.2} right panel, blue curve). The potential scattering  explicitly breaks the PH symmetry. It can be accounted by replacement of $\delta_0$ by 
$\tilde \delta_0$$=$$\delta_0$$+$$\delta_P$, $\delta_P$$\ll $$ \delta_0$.
As a result, the finite $G_{12}$ and thermopower arises. The results of calculations obtained from the zero-current conditions for SU(4) quantum impurity are illustrated on Fig. \ref{f.2}. First, for the case of singly occupied quantum impurity $m$$=$$1$ (Fig. \ref{f.2} left panel) $\delta_0$$=$$\pi/4$ and the PH symmetry is explicitly broken. At zero potential scattering inelastic effects associated with the finite bias voltage vanish and the thermoelectric power is completely defined by elastic processes. The effect of potential scattering is two-fold: i) it detunes $G_{12}$ from it's maximal value and ii) it results in finite-temperature inelastic corrections to the conductance (see the Fig. \ref{f.2} insert). For the PHS case of double occupation $m$$=$$2$ (Fig. \ref{f.2} right panel) $\delta_0$$=$$\pi/2$. Therefore the finite potential scattering results in finite $G_{12}$ and differential thermopower $S$ is proportional to $\sin 2\delta_P$. 
{\color{black} Crossover between $m$$=$$2$ SU(4) and $m$$=$$1$
SU(2) has been studied recently experimentally \cite{Ferrier_PRL_2017}.} 
Note that the current across the dot symmetrically coupled to the leads  contains only odd powers of the voltage both for the PHS 
$m$$=$$2$ and the PH-non-symmetric (PHN) $m$$=$$1$ cases.
\footnote{It is instructive to re-write conductance in (\ref{elecondu}) as $G(T)$$=$$G(0)$$($$1$$-$$c_T$$(\pi T/T_K)^2)$,
$G$$($$0)$$=$$G_0$$\sin^2x$ and the FL constant 
$c_T$$=$$($$N$$+$$1$$)$$/$$($$3$$($$N$$-$$1$$)$$)$$($$2$$\cos$$ 2x$$)$$/$$($$\cos$$ 2x$$-$$1$$)$.
Potential scattering results in the replacement $x$$=$$\delta_0$$\to$$\delta_0$$+$$\delta_P$
and renormalizes both $G$$($$0)$ and $c_T$ \cite{GP_Review_2005}.}

{\it Seebeck effect}. In the limit $T$$\rightarrow $$0$ the differential thermopower is given by \footnote{This relation is also know as the Mott-Cutler formula \cite{Costi_Zlatic_PRB(81)_2010}.}
\begin{eqnarray}\label{scaling}
S(T)=-2eL_0T\cot\tilde\delta_0 \left. \frac{\partial \tilde\delta(\varepsilon)}{\partial \varepsilon}\right|_{\varepsilon=0} =-\frac{\pi\gamma T}{e}\cot \tilde\delta_0,\;\;\;\;\;\;
\end{eqnarray}
where $L_0$$=$$\pi^2$$/$$(3e^2)$ is the Lorentz number and {\color{black} 
$\gamma$$=$$2$$\pi$$\alpha_1/3$$\sim$$1$$/T_K$}
in accordance with the FL theory \cite{Hewson}. 

The thermopower {\it measurements} \cite{Scheibner_PRL(95)_2005} 
refer however {\it not} to the differential Seebeck effects 
(see Fig. \ref{f.1}). Since there were no independent measurements of $T_L$ and $T_R$, the temperature drop
was estimated from the Joule heat. It appeared that the $\Delta T$ was finite and {\it not}
fulfilling the condition $|\Delta T|$$ \ll $$T_R$. To demonstrate the significance of non-linear effects associated with finite temperature drop we show on Fig. \ref{f.3} the thermopower of SU(4) model 
computed by two different methods: i) the dashed blue line stands for the differential thermopower 
$S(T)$$=$$G_{12}$$/$$G$ where $G_{12}$ is obtained at 
zero voltage drop while $G$ is calculated at equal temperatures of the leads; ii) the solid lines corresponds to $S$$=$$-$$\Delta$$V$$/$$\Delta $$T$ resembling the experimental situation in \cite{Scheibner_PRL(95)_2005} : the temperature drop is fixed 
$\Delta$$T$$/$$T_K$$=0.05$ (red) $0.025$ (green) $0.01$ (orange) and the thermo-voltage is obtained from the zero current condition. As one can see, at small reference temperatures "finite $\Delta T$" thermopower always {\it overshoots}
the differential $S$. The effect is more pronounced in PHS regime \footnote{The offset can be easily understood on trivial example of SU(4) $m$$=$$1$ and $\delta_P$$=$$0$. In that case the inelastic contribution to the current vanish and the non-linear effect is $\propto$ $(\Delta T)^2$. Therefore, the offset is linear in $\Delta T$ and can be used as a measure of the temperature drop.}. This observation can explain the thermo-voltage offset observed in the experiment \cite{Scheibner_PRL(95)_2005} in the Kondo limit of SU(2) quantum impurity (PHS regime). According to our calculations this offset is associated with a non-linear $\Delta T$ dependence of the current at low reference temperatures (see Fig. \ref{f.3} inserts).  
We suggest to check this statement experimentally by performing Seebeck effect measurements 
varying the temperature in the "hot" lead.

{\it Coupling asymmetry.} The effect of coupling asymmetry $t_L$$\neq $$t_R$ in (\ref{HamAnd}) manifests itself in the following way:
for broken  PH-symmetry case it results in an asymmetric I-V curve due to a
contribution to the current quadratic in voltage  which, in turn, depends linearly on the coupling asymmetry. For both PHS and PHN cases
the coupling asymmetry results in {\color{black} i) renormalization of
the elastic contribution to the charge current (see \cite{Karki_MK_Supplement} Eq. S5); ii) renormalization of the Kondo temperature due to tunneling rates asymmetry \cite{Karki_MK_Supplement};
and} iii) renormalization of the coefficient in front of the term cubic in voltage. The magnitude of current is suppressed by the coupling asymmetry. Besides, it also affects the thermo-current. 
However, this effect is proportional to $\Delta$$V$$\cdot \Delta T$ and therefore beyond the linear response theory (see the Supplemental Materials  \cite{Karki_MK_Supplement} for details).  

{\it Peltier effect.}
In order to compute other thermo-electric coefficients, e.g. Peltier effect, one needs to define and compute the heat current. To proceed with full fledged Keldysh calculations one can e.g. deal with the Luttinger "gravitational potential" approach 
\cite{Luttinger_PR_(135)_1964,Shastry_RPP_(72)_2009,Eich_PRB_(90)_2014}. 
Such a theory would access the 
effects non-linear in $\Delta T$ \cite{Karki_MK_TBP}. In the linear response theory the Peltier coefficient $\Pi$$($$T)$ can be calculated using the transport integrals method (see, e.g. \cite{Costi_Zlatic_PRB(81)_2010}
for details) based on calculation of different momenta of the single-particle lifetime
$\tau$$($$\epsilon$$,$$T)$ (see the Supplemental materials \cite{Karki_MK_Supplement}) and is related to the thermopower by the Onsager's relation $\Pi(T)$$=$$S$$($$T$$)$$\cdot T$.
\footnote{
The Lorenz number $L_0(T)$ for given $T$ connects the
thermal $K_e(T)$ and electrical $G(T)$ conductances: $L_0$$($$T$$)$$=$$K_e$$($$T$$)$$/$$T$$G$$(T)$ \cite{Karki_MK_Supplement}. 
The figure of merit (neglecting the phonon contribution) is defined by
$ZT$$=$$S^2(T)$$/$$L_0(T)$. The highest $ZT$ is achieved in the PHN regime at $\delta_P$$=$$0$ and symmetric dot-leads coupling.}

\noindent {\it Summary.}
\noindent The full fledged theory based on Keldysh out-of equilibrium calculations of the electric current
is constructed for the SU(N) Kondo quantum impurity subject to a finite bias voltage and a finite temperature drop. The transport coefficients: conductance $G$, thermoelectric coefficient $G_{12}$ and thermopower $S$ are computed under condition of zero-current state for the strong-coupling regime of the quantum impurity. It is shown that pronounced non-linear effects in temperature drop influence the transport coefficients at the low-temperature limit. These effects are likely sufficient to resolve 
the experimental puzzle of the thermo-transport through the Kondo impurity at the strong coupling.\\
\noindent {\it Acknowledgement.} We acknowledge fruitful discussions with Jan von Delft, Laurens W. Molenkamp and Christophe Mora. We are grateful to Yu. Galperin and S. Ludwig for careful reading the manuscript, comments and suggestions. {\color{black} This work was finalized at Aspen Center for Physics, which is supported by National Science Foundation grant
PHY-1607611 and was partially supported (MK) by a grant from the Simons Foundation.}\\
\noindent {\it Note added.} While completing this {\color{black} work, a paper} \cite{Lopez_arXiv_2017}
appeared where the significance of non-linear effects in temperature drop on the temperature-driven current through SU(2) quantum impurity has been reported.


\begin{thebibliography}{65}%
\makeatletter
\providecommand \@ifxundefined [1]{%
 \@ifx{#1\undefined}
}%
\providecommand \@ifnum [1]{%
 \ifnum #1\expandafter \@firstoftwo
 \else \expandafter \@secondoftwo
 \fi
}%
\providecommand \@ifx [1]{%
 \ifx #1\expandafter \@firstoftwo
 \else \expandafter \@secondoftwo
 \fi
}%
\providecommand \natexlab [1]{#1}%
\providecommand \enquote  [1]{``#1''}%
\providecommand \bibnamefont  [1]{#1}%
\providecommand \bibfnamefont [1]{#1}%
\providecommand \citenamefont [1]{#1}%
\providecommand \href@noop [0]{\@secondoftwo}%
\providecommand \href [0]{\begingroup \@sanitize@url \@href}%
\providecommand \@href[1]{\@@startlink{#1}\@@href}%
\providecommand \@@href[1]{\endgroup#1\@@endlink}%
\providecommand \@sanitize@url [0]{\catcode `\\12\catcode `\$12\catcode
  `\&12\catcode `\#12\catcode `\^12\catcode `\_12\catcode `\%12\relax}%
\providecommand \@@startlink[1]{}%
\providecommand \@@endlink[0]{}%
\providecommand \url  [0]{\begingroup\@sanitize@url \@url }%
\providecommand \@url [1]{\endgroup\@href {#1}{\urlprefix }}%
\providecommand \urlprefix  [0]{URL }%
\providecommand \Eprint [0]{\href }%
\providecommand \doibase [0]{http://dx.doi.org/}%
\providecommand \selectlanguage [0]{\@gobble}%
\providecommand \bibinfo  [0]{\@secondoftwo}%
\providecommand \bibfield  [0]{\@secondoftwo}%
\providecommand \translation [1]{[#1]}%
\providecommand \BibitemOpen [0]{}%
\providecommand \bibitemStop [0]{}%
\providecommand \bibitemNoStop [0]{.\EOS\space}%
\providecommand \EOS [0]{\spacefactor3000\relax}%
\providecommand \BibitemShut  [1]{\csname bibitem#1\endcsname}%
\let\auto@bib@innerbib\@empty
\bibitem [{\citenamefont {Jezouin}\ \emph {et~al.}(2016)\citenamefont
  {Jezouin}, \citenamefont {Iftikhar}, \citenamefont {Anthore}, \citenamefont
  {Parmentier}, \citenamefont {Gennser}, \citenamefont {Cavanna}, \citenamefont
  {Ouerghi}, \citenamefont {Levkivskyi}, \citenamefont {Idrisov}, \citenamefont
  {Sukhorukov}, \citenamefont {Glazman},\ and\ \citenamefont
  {Pierre}}]{Pierre_NAT(536)_2016}%
  \BibitemOpen
  \bibfield  {author} {\bibinfo {author} {\bibfnamefont {S.}~\bibnamefont
  {Jezouin}}, \bibinfo {author} {\bibfnamefont {Z.}~\bibnamefont {Iftikhar}},
  \bibinfo {author} {\bibfnamefont {A.}~\bibnamefont {Anthore}}, \bibinfo
  {author} {\bibfnamefont {F.~D.}\ \bibnamefont {Parmentier}}, \bibinfo
  {author} {\bibfnamefont {U.}~\bibnamefont {Gennser}}, \bibinfo {author}
  {\bibfnamefont {A.}~\bibnamefont {Cavanna}}, \bibinfo {author} {\bibfnamefont
  {A.}~\bibnamefont {Ouerghi}}, \bibinfo {author} {\bibfnamefont {I.~P.}\
  \bibnamefont {Levkivskyi}}, \bibinfo {author} {\bibfnamefont
  {E.}~\bibnamefont {Idrisov}}, \bibinfo {author} {\bibfnamefont {E.~V.}\
  \bibnamefont {Sukhorukov}}, \bibinfo {author} {\bibfnamefont {L.~I.}\
  \bibnamefont {Glazman}}, \ and\ \bibinfo {author} {\bibfnamefont
  {F.}~\bibnamefont {Pierre}},\ }\href@noop {} {\bibfield  {journal} {\bibinfo
  {journal} {Nature}\ }\textbf {\bibinfo {volume} {536}},\ \bibinfo {pages}
  {60} (\bibinfo {year} {2016})}\BibitemShut {NoStop}%
\bibitem [{\citenamefont {Iftikhar}\ \emph {et~al.}(2015)\citenamefont
  {Iftikhar}, \citenamefont {Jezouin}, \citenamefont {Anthore}, \citenamefont
  {Gennser}, \citenamefont {Parmentier}, \citenamefont {Cavanna},\ and\
  \citenamefont {Pierre}}]{Pierre_NAT(526)_2015}%
  \BibitemOpen
  \bibfield  {author} {\bibinfo {author} {\bibfnamefont {Z.}~\bibnamefont
  {Iftikhar}}, \bibinfo {author} {\bibfnamefont {S.}~\bibnamefont {Jezouin}},
  \bibinfo {author} {\bibfnamefont {A.}~\bibnamefont {Anthore}}, \bibinfo
  {author} {\bibfnamefont {U.}~\bibnamefont {Gennser}}, \bibinfo {author}
  {\bibfnamefont {F.~D.}\ \bibnamefont {Parmentier}}, \bibinfo {author}
  {\bibfnamefont {A.}~\bibnamefont {Cavanna}}, \ and\ \bibinfo {author}
  {\bibfnamefont {F.}~\bibnamefont {Pierre}},\ }\href@noop {} {\bibfield
  {journal} {\bibinfo  {journal} {Nature}\ }\textbf {\bibinfo {volume} {526}},\
  \bibinfo {pages} {233} (\bibinfo {year} {2015})}\BibitemShut {NoStop}%
\bibitem [{\citenamefont {Jezouin}\ \emph {et~al.}(2013)\citenamefont
  {Jezouin}, \citenamefont {Parmentier}, \citenamefont {Anthore}, \citenamefont
  {Gennser}, \citenamefont {Cavanna}, \citenamefont {Jin},\ and\ \citenamefont
  {Pierre}}]{Pierre_SCI(342)_2013}%
  \BibitemOpen
  \bibfield  {author} {\bibinfo {author} {\bibfnamefont {S.}~\bibnamefont
  {Jezouin}}, \bibinfo {author} {\bibfnamefont {F.~D.}\ \bibnamefont
  {Parmentier}}, \bibinfo {author} {\bibfnamefont {A.}~\bibnamefont {Anthore}},
  \bibinfo {author} {\bibfnamefont {U.}~\bibnamefont {Gennser}}, \bibinfo
  {author} {\bibfnamefont {A.}~\bibnamefont {Cavanna}}, \bibinfo {author}
  {\bibfnamefont {Y.}~\bibnamefont {Jin}}, \ and\ \bibinfo {author}
  {\bibfnamefont {F.}~\bibnamefont {Pierre}},\ }\href@noop {} {\bibfield
  {journal} {\bibinfo  {journal} {Science}\ }\textbf {\bibinfo {volume}
  {342}},\ \bibinfo {pages} {601} (\bibinfo {year} {2013})}\BibitemShut
  {NoStop}%
{\color{black}\bibitem [{\citenamefont {Benenti}\ \emph {et~al.}(2017)\citenamefont
  {Benenti}, \citenamefont {Casati}, \citenamefont {Saito},\ and\ \citenamefont
  {Whitney}}]{Thermo_review}%
  \BibitemOpen
  \bibfield  {author} {\bibinfo {author} {\bibfnamefont {G.}~\bibnamefont
  {Benenti}}, \bibinfo {author} {\bibfnamefont {G.}~\bibnamefont {Casati}},
  \bibinfo {author} {\bibfnamefont {K.}~\bibnamefont {Saito}}, \ and\ \bibinfo
  {author} {\bibfnamefont {R.~S.}\ \bibnamefont {Whitney}},\ }\href@noop {}
  {\bibfield  {journal} {\bibinfo  {journal} {Physics Reports}\ }\textbf
  {\bibinfo {volume} {694}},\ \bibinfo {pages} {1} (\bibinfo {year}
  {2017})}}\BibitemShut {NoStop}%
\bibitem [{\citenamefont {Blanter}\ and\ \citenamefont
  {Nazarov}(2009)}]{Blanter_Nazarov}%
  \BibitemOpen
  \bibfield  {author} {\bibinfo {author} {\bibfnamefont {Y.~M.}\ \bibnamefont
  {Blanter}}\ and\ \bibinfo {author} {\bibfnamefont {Y.~V.}\ \bibnamefont
  {Nazarov}},\ }\href@noop {} {\emph {\bibinfo {title} {Quantum Transport:
  Introduction to Nanoscience}}}\ (\bibinfo  {publisher} {Cambridge University
  Press, Cambridge},\ \bibinfo {year} {2009})\BibitemShut {NoStop}%
\bibitem [{\citenamefont {Molenkamp}\ \emph {et~al.}(1990)\citenamefont
  {Molenkamp}, \citenamefont {vanHouten}, \citenamefont {Beenakker},
  \citenamefont {Eppenga},\ and\ \citenamefont {Foxon}}]{Molenkamp_1990}%
  \BibitemOpen
  \bibfield  {author} {\bibinfo {author} {\bibfnamefont {L.~W.}\ \bibnamefont
  {Molenkamp}}, \bibinfo {author} {\bibfnamefont {H.}~\bibnamefont
  {vanHouten}}, \bibinfo {author} {\bibfnamefont {C.~W.~J.}\ \bibnamefont
  {Beenakker}}, \bibinfo {author} {\bibfnamefont {R.}~\bibnamefont {Eppenga}},
  \ and\ \bibinfo {author} {\bibfnamefont {C.~T.}\ \bibnamefont {Foxon}},\
  }\href@noop {} {\bibfield  {journal} {\bibinfo  {journal} {Phys. Rev. Lett}\
  }\textbf {\bibinfo {volume} {65}},\ \bibinfo {pages} {1052} (\bibinfo {year}
  {1990})}\BibitemShut {NoStop}%
\bibitem [{\citenamefont {vanHouten}\ \emph {et~al.}(1992)\citenamefont
  {vanHouten}, \citenamefont {Molenkamp}, \citenamefont {Beenakker},\ and\
  \citenamefont {Foxon}}]{Molenkamp_1992}%
  \BibitemOpen
  \bibfield  {author} {\bibinfo {author} {\bibfnamefont {H.}~\bibnamefont
  {vanHouten}}, \bibinfo {author} {\bibfnamefont {L.~W.}\ \bibnamefont
  {Molenkamp}}, \bibinfo {author} {\bibfnamefont {C.~W.~J.}\ \bibnamefont
  {Beenakker}}, \ and\ \bibinfo {author} {\bibfnamefont {C.~T.}\ \bibnamefont
  {Foxon}},\ }\href@noop {} {\bibfield  {journal} {\bibinfo  {journal}
  {Semicond. Sci. Technol.}\ }\textbf {\bibinfo {volume} {7}},\ \bibinfo
  {pages} {B215} (\bibinfo {year} {1992})}\BibitemShut {NoStop}%
\bibitem [{\citenamefont {Scheibner}\ \emph {et~al.}(2005)\citenamefont
  {Scheibner}, \citenamefont {Buhmann}, \citenamefont {Reuter}, \citenamefont
  {Kiselev},\ and\ \citenamefont {Molenkamp}}]{Scheibner_PRL(95)_2005}%
  \BibitemOpen
  \bibfield  {author} {\bibinfo {author} {\bibfnamefont {R.}~\bibnamefont
  {Scheibner}}, \bibinfo {author} {\bibfnamefont {H.}~\bibnamefont {Buhmann}},
  \bibinfo {author} {\bibfnamefont {D.}~\bibnamefont {Reuter}}, \bibinfo
  {author} {\bibfnamefont {M.~N.}\ \bibnamefont {Kiselev}}, \ and\ \bibinfo
  {author} {\bibfnamefont {L.~W.}\ \bibnamefont {Molenkamp}},\ }\href@noop {}
  {\bibfield  {journal} {\bibinfo  {journal} {Phys. Rev. Lett.}\ }\textbf
  {\bibinfo {volume} {95}},\ \bibinfo {pages} {176602} (\bibinfo {year}
  {2005})}\BibitemShut {NoStop}%
\bibitem [{\citenamefont {Zlatic}\ and\ \citenamefont
  {Monnier}(2014)}]{Zlatic}%
  \BibitemOpen
  \bibfield  {author} {\bibinfo {author} {\bibfnamefont {V.}~\bibnamefont
  {Zlatic}}\ and\ \bibinfo {author} {\bibfnamefont {R.}~\bibnamefont
  {Monnier}},\ }\href@noop {} {\emph {\bibinfo {title} {Modern Theory of
  Thermoelectricity}}}\ (\bibinfo  {publisher} {Oxford University Press},\
  \bibinfo {year} {2014})\BibitemShut {NoStop}%
\bibitem [{\citenamefont {Nozi{\'e}res}(1974)}]{Nozieres}%
  \BibitemOpen
  \bibfield  {author} {\bibinfo {author} {\bibfnamefont {P.}~\bibnamefont
  {Nozi{\'e}res}},\ }\href@noop {} {\bibfield  {journal} {\bibinfo  {journal}
  {J. Low Temp. Phys.}\ }\textbf {\bibinfo {volume} {17}} (\bibinfo {year}
  {1974})}\BibitemShut {NoStop}%
\bibitem [{\citenamefont {Nozieres}\ and\ \citenamefont
  {Blandin}(1980)}]{Nozieres_Blandin_JPhys_1980}%
  \BibitemOpen
  \bibfield  {author} {\bibinfo {author} {\bibfnamefont {P.}~\bibnamefont
  {Nozieres}}\ and\ \bibinfo {author} {\bibfnamefont {A.}~\bibnamefont
  {Blandin}},\ }\href@noop {} {\bibfield  {journal} {\bibinfo  {journal} {J.
  Phys}\ }\textbf {\bibinfo {volume} {41}},\ \bibinfo {pages} {193} (\bibinfo
  {year} {1980})}\BibitemShut {NoStop}%
\bibitem [{\citenamefont {Hewson}(1993)}]{Hewson}%
  \BibitemOpen
  \bibfield  {author} {\bibinfo {author} {\bibfnamefont {A.}~\bibnamefont
  {Hewson}},\ }\href@noop {} {\emph {\bibinfo {title} {The Kondo Problem to
  Heavy Fermions}}}\ (\bibinfo  {publisher} {Cambridge University Press,
  Cambridge},\ \bibinfo {year} {1993})\BibitemShut {NoStop}%
\bibitem [{\citenamefont {Goldhaber-Gordon}\ \emph {et~al.}(1998)\citenamefont
  {Goldhaber-Gordon}, \citenamefont {Shtrikman}, \citenamefont {Mahalu},
  \citenamefont {Abusch-Magder}, \citenamefont {Meirav},\ and\ \citenamefont
  {Kastner}}]{Goldhaber_nat(391)_1998}%
  \BibitemOpen
  \bibfield  {author} {\bibinfo {author} {\bibfnamefont {D.}~\bibnamefont
  {Goldhaber-Gordon}}, \bibinfo {author} {\bibfnamefont {H.}~\bibnamefont
  {Shtrikman}}, \bibinfo {author} {\bibfnamefont {D.}~\bibnamefont {Mahalu}},
  \bibinfo {author} {\bibfnamefont {D.}~\bibnamefont {Abusch-Magder}}, \bibinfo
  {author} {\bibfnamefont {U.}~\bibnamefont {Meirav}}, \ and\ \bibinfo {author}
  {\bibfnamefont {M.~A.}\ \bibnamefont {Kastner}},\ }\href@noop {} {\bibfield
  {journal} {\bibinfo  {journal} {Nature}\ }\textbf {\bibinfo {volume} {391}},\
  \bibinfo {pages} {156} (\bibinfo {year} {1998})}\BibitemShut {NoStop}%
\bibitem [{\citenamefont {Cronenwett}\ \emph {et~al.}(1998)\citenamefont
  {Cronenwett}, \citenamefont {Oosterkamp},\ and\ \citenamefont
  {Kouwenhoven}}]{Cronewett_SCI(281)_1998}%
  \BibitemOpen
  \bibfield  {author} {\bibinfo {author} {\bibfnamefont {S.~M.}\ \bibnamefont
  {Cronenwett}}, \bibinfo {author} {\bibfnamefont {T.~H.}\ \bibnamefont
  {Oosterkamp}}, \ and\ \bibinfo {author} {\bibfnamefont {L.~P.}\ \bibnamefont
  {Kouwenhoven}},\ }\href@noop {} {\bibfield  {journal} {\bibinfo  {journal}
  {Science}\ }\textbf {\bibinfo {volume} {281}},\ \bibinfo {pages} {540}
  (\bibinfo {year} {1998})}\BibitemShut {NoStop}%
\bibitem [{\citenamefont {Nyg{\aa}ard}\ \emph {et~al.}(2000)\citenamefont
  {Nyg{\aa}ard}, \citenamefont {Cobden},\ and\ \citenamefont
  {Lindelof}}]{Jesper_NAT(408)_2000}%
  \BibitemOpen
  \bibfield  {author} {\bibinfo {author} {\bibfnamefont {J.}~\bibnamefont
  {Nyg{\aa}ard}}, \bibinfo {author} {\bibfnamefont {D.~H.}\ \bibnamefont
  {Cobden}}, \ and\ \bibinfo {author} {\bibfnamefont {P.~E.}\ \bibnamefont
  {Lindelof}},\ }\href@noop {} {\bibfield  {journal} {\bibinfo  {journal}
  {Nature}\ }\textbf {\bibinfo {volume} {408}},\ \bibinfo {pages} {342}
  (\bibinfo {year} {2000})}\BibitemShut {NoStop}%
\bibitem [{\citenamefont {Schr\"oer}\ \emph {et~al.}(2006)\citenamefont
  {Schr\"oer}, \citenamefont {H\"uttel}, \citenamefont {Eberl}, \citenamefont
  {Ludwig}, \citenamefont {Kiselev},\ and\ \citenamefont
  {Altshuler}}]{Ludwig_2006}%
  \BibitemOpen
  \bibfield  {author} {\bibinfo {author} {\bibfnamefont {D.~M.}\ \bibnamefont
  {Schr\"oer}}, \bibinfo {author} {\bibfnamefont {A.~K.}\ \bibnamefont
  {H\"uttel}}, \bibinfo {author} {\bibfnamefont {K.}~\bibnamefont {Eberl}},
  \bibinfo {author} {\bibfnamefont {S.}~\bibnamefont {Ludwig}}, \bibinfo
  {author} {\bibfnamefont {M.~N.}\ \bibnamefont {Kiselev}}, \ and\ \bibinfo
  {author} {\bibfnamefont {B.~L.}\ \bibnamefont {Altshuler}},\ }\href@noop {}
  {\bibfield  {journal} {\bibinfo  {journal} {Phys. Rev. B}\ }\textbf {\bibinfo
  {volume} {74}},\ \bibinfo {pages} {233301} (\bibinfo {year}
  {2006})}\BibitemShut {NoStop}%
\bibitem [{\citenamefont {Affleck}\ and\ \citenamefont
  {Ludwig}(1993)}]{Affleck_Lud_PRB(48)_1993}%
  \BibitemOpen
  \bibfield  {author} {\bibinfo {author} {\bibfnamefont {I.}~\bibnamefont
  {Affleck}}\ and\ \bibinfo {author} {\bibfnamefont {A.~W.~W.}\ \bibnamefont
  {Ludwig}},\ }\href@noop {} {\bibfield  {journal} {\bibinfo  {journal} {Phys.
  Rev. B}\ }\textbf {\bibinfo {volume} {48}},\ \bibinfo {pages} {7297}
  (\bibinfo {year} {1993})}\BibitemShut {NoStop}%
\bibitem [{\citenamefont {Mora}(2009)}]{Mora_PRB_(80)_2009}%
  \BibitemOpen
  \bibfield  {author} {\bibinfo {author} {\bibfnamefont {C.}~\bibnamefont
  {Mora}},\ }\href@noop {} {\bibfield  {journal} {\bibinfo  {journal} {Phys.
  Rev. B}\ }\textbf {\bibinfo {volume} {80}},\ \bibinfo {pages} {125304}
  (\bibinfo {year} {2009})}\BibitemShut {NoStop}%
\bibitem [{\citenamefont {Vitushinsky}\ \emph {et~al.}(2008)\citenamefont
  {Vitushinsky}, \citenamefont {Clerk},\ and\ \citenamefont
  {LeHur}}]{Vitushinsky_Clerk_Hur_PRL_(100)_2008}%
  \BibitemOpen
  \bibfield  {author} {\bibinfo {author} {\bibfnamefont {P.}~\bibnamefont
  {Vitushinsky}}, \bibinfo {author} {\bibfnamefont {A.~A.}\ \bibnamefont
  {Clerk}}, \ and\ \bibinfo {author} {\bibfnamefont {K.}~\bibnamefont
  {LeHur}},\ }\href@noop {} {\bibfield  {journal} {\bibinfo  {journal} {Phys.
  Rev. Lett.}\ }\textbf {\bibinfo {volume} {100}},\ \bibinfo {pages} {036603}
  (\bibinfo {year} {2008})}\BibitemShut {NoStop}%
\bibitem [{\citenamefont {Mora}\ \emph {et~al.}(2008)\citenamefont {Mora},
  \citenamefont {Leyronas},\ and\ \citenamefont
  {Regnault}}]{Mora_Leyronas_Regnault_PRL_(100)_2008}%
  \BibitemOpen
  \bibfield  {author} {\bibinfo {author} {\bibfnamefont {C.}~\bibnamefont
  {Mora}}, \bibinfo {author} {\bibfnamefont {X.}~\bibnamefont {Leyronas}}, \
  and\ \bibinfo {author} {\bibfnamefont {N.}~\bibnamefont {Regnault}},\
  }\href@noop {} {\bibfield  {journal} {\bibinfo  {journal} {Phys. Rev. Lett.}\
  }\textbf {\bibinfo {volume} {100}},\ \bibinfo {pages} {036604} (\bibinfo
  {year} {2008})}\BibitemShut {NoStop}%
\bibitem [{\citenamefont {Mora}\ \emph {et~al.}(2009)\citenamefont {Mora},
  \citenamefont {Vitushinsky}, \citenamefont {Leyronas}, \citenamefont
  {Clerk},\ and\ \citenamefont {LeHur}}]{Mora_Clerk_Hur_PRB(80)_2009}%
  \BibitemOpen
  \bibfield  {author} {\bibinfo {author} {\bibfnamefont {C.}~\bibnamefont
  {Mora}}, \bibinfo {author} {\bibfnamefont {P.}~\bibnamefont {Vitushinsky}},
  \bibinfo {author} {\bibfnamefont {X.}~\bibnamefont {Leyronas}}, \bibinfo
  {author} {\bibfnamefont {A.~A.}\ \bibnamefont {Clerk}}, \ and\ \bibinfo
  {author} {\bibfnamefont {K.}~\bibnamefont {LeHur}},\ }\href@noop {}
  {\bibfield  {journal} {\bibinfo  {journal} {Phys. Rev. B}\ }\textbf {\bibinfo
  {volume} {80}},\ \bibinfo {pages} {155322} (\bibinfo {year}
  {2009})}\BibitemShut {NoStop}%
\bibitem [{\citenamefont {Mora}\ \emph {et~al.}(2015)\citenamefont {Mora},
  \citenamefont {Moca}, \citenamefont {von Delft},\ and\ \citenamefont
  {Zar{\'a}nd}}]{Mora_Moca_Delft_Zarand_PRB(92)_2015}%
  \BibitemOpen
  \bibfield  {author} {\bibinfo {author} {\bibfnamefont {C.}~\bibnamefont
  {Mora}}, \bibinfo {author} {\bibfnamefont {C.~P.}\ \bibnamefont {Moca}},
  \bibinfo {author} {\bibfnamefont {J.}~\bibnamefont {von Delft}}, \ and\
  \bibinfo {author} {\bibfnamefont {G.}~\bibnamefont {Zar{\'a}nd}},\
  }\href@noop {} {\bibfield  {journal} {\bibinfo  {journal} {Phys. Rev. B}\
  }\textbf {\bibinfo {volume} {92}},\ \bibinfo {pages} {075120} (\bibinfo
  {year} {2015})}\BibitemShut {NoStop}%
\bibitem [{\citenamefont {LeHur}\ \emph {et~al.}(2007)\citenamefont {LeHur},
  \citenamefont {Simon},\ and\ \citenamefont
  {Loss}}]{Hur_Simon_Loss_PRB(75)_2007}%
  \BibitemOpen
  \bibfield  {author} {\bibinfo {author} {\bibfnamefont {K.}~\bibnamefont
  {LeHur}}, \bibinfo {author} {\bibfnamefont {P.}~\bibnamefont {Simon}}, \ and\
  \bibinfo {author} {\bibfnamefont {D.}~\bibnamefont {Loss}},\ }\href@noop {}
  {\bibfield  {journal} {\bibinfo  {journal} {Phys. Rev. B}\ }\textbf {\bibinfo
  {volume} {75}},\ \bibinfo {pages} {035332} (\bibinfo {year}
  {2007})}\BibitemShut {NoStop}%
\bibitem [{\citenamefont {H\"{o}rig}\ \emph {et~al.}(2014)\citenamefont
  {H\"{o}rig}, \citenamefont {Mora},\ and\ \citenamefont
  {Schuricht}}]{Mora_Schuricht_(89)_2014}%
  \BibitemOpen
  \bibfield  {author} {\bibinfo {author} {\bibfnamefont {C.~B.~M.}\
  \bibnamefont {H\"{o}rig}}, \bibinfo {author} {\bibfnamefont {C.}~\bibnamefont
  {Mora}}, \ and\ \bibinfo {author} {\bibfnamefont {D.}~\bibnamefont
  {Schuricht}},\ }\href@noop {} {\bibfield  {journal} {\bibinfo  {journal}
  {Phys. Rev. B}\ }\textbf {\bibinfo {volume} {89}},\ \bibinfo {pages} {165411}
  (\bibinfo {year} {2014})}\BibitemShut {NoStop}%
\bibitem [{\citenamefont {Hanl}\ \emph {et~al.}(2014)\citenamefont {Hanl},
  \citenamefont {Weichselbaum}, \citenamefont {von Delft},\ and\ \citenamefont
  {Kiselev}}]{HWDK_PRB_(89)_2014}%
  \BibitemOpen
  \bibfield  {author} {\bibinfo {author} {\bibfnamefont {M.}~\bibnamefont
  {Hanl}}, \bibinfo {author} {\bibfnamefont {A.}~\bibnamefont {Weichselbaum}},
  \bibinfo {author} {\bibfnamefont {J.}~\bibnamefont {von Delft}}, \ and\
  \bibinfo {author} {\bibfnamefont {M.}~\bibnamefont {Kiselev}},\ }\href@noop
  {} {\bibfield  {journal} {\bibinfo  {journal} {Phys. Rev. B}\ }\textbf
  {\bibinfo {volume} {89}},\ \bibinfo {pages} {195131} (\bibinfo {year}
  {2014})}\BibitemShut {NoStop}%
\bibitem [{\citenamefont {Rejec}\ and\ \citenamefont
  {Meir}(2006)}]{Meir_NAT_(442)_2006}%
  \BibitemOpen
  \bibfield  {author} {\bibinfo {author} {\bibfnamefont {T.}~\bibnamefont
  {Rejec}}\ and\ \bibinfo {author} {\bibfnamefont {Y.}~\bibnamefont {Meir}},\
  }\href@noop {} {\bibfield  {journal} {\bibinfo  {journal} {Nature}\ }\textbf
  {\bibinfo {volume} {442}},\ \bibinfo {pages} {900} (\bibinfo {year}
  {2006})}\BibitemShut {NoStop}%
\bibitem [{\citenamefont {Bauer}\ \emph
  {et~al.}(2013{\natexlab{a}})\citenamefont {Bauer}, \citenamefont {Heyder},
  \citenamefont {Schubert}, \citenamefont {Borowsky}, \citenamefont {Taubert},
  \citenamefont {Bruognolo}, \citenamefont {Schuh}, \citenamefont
  {Wegscheider}, \citenamefont {von Delft},\ and\ \citenamefont
  {Ludwig}}]{JVD_NAT_(501)_2013}%
  \BibitemOpen
  \bibfield  {author} {\bibinfo {author} {\bibfnamefont {F.}~\bibnamefont
  {Bauer}}, \bibinfo {author} {\bibfnamefont {J.}~\bibnamefont {Heyder}},
  \bibinfo {author} {\bibfnamefont {E.}~\bibnamefont {Schubert}}, \bibinfo
  {author} {\bibfnamefont {D.}~\bibnamefont {Borowsky}}, \bibinfo {author}
  {\bibfnamefont {D.}~\bibnamefont {Taubert}}, \bibinfo {author} {\bibfnamefont
  {B.}~\bibnamefont {Bruognolo}}, \bibinfo {author} {\bibfnamefont
  {D.}~\bibnamefont {Schuh}}, \bibinfo {author} {\bibfnamefont
  {W.}~\bibnamefont {Wegscheider}}, \bibinfo {author} {\bibfnamefont
  {J.}~\bibnamefont {von Delft}}, \ and\ \bibinfo {author} {\bibfnamefont
  {S.}~\bibnamefont {Ludwig}},\ }\href@noop {} {\bibfield  {journal} {\bibinfo
  {journal} {Nature}\ }\textbf {\bibinfo {volume} {501}},\ \bibinfo {pages}
  {73} (\bibinfo {year} {2013}{\natexlab{a}})}\BibitemShut {NoStop}%
\bibitem [{\citenamefont {Jarillo-Herrero}\ \emph {et~al.}(2005)\citenamefont
  {Jarillo-Herrero}, \citenamefont {Kong}, \citenamefont {van~der Zant},
  \citenamefont {Dekker}, \citenamefont {Kouwenhoven},\ and\ \citenamefont
  {Franceschi}}]{Jarillo-Herrero_Nat(434)_2005}%
  \BibitemOpen
  \bibfield  {author} {\bibinfo {author} {\bibfnamefont {P.}~\bibnamefont
  {Jarillo-Herrero}}, \bibinfo {author} {\bibfnamefont {J.}~\bibnamefont
  {Kong}}, \bibinfo {author} {\bibfnamefont {H.~S.}\ \bibnamefont {van~der
  Zant}}, \bibinfo {author} {\bibfnamefont {C.}~\bibnamefont {Dekker}},
  \bibinfo {author} {\bibfnamefont {L.~P.}\ \bibnamefont {Kouwenhoven}}, \ and\
  \bibinfo {author} {\bibfnamefont {S.~D.}\ \bibnamefont {Franceschi}},\
  }\href@noop {} {\bibfield  {journal} {\bibinfo  {journal} {Nature}\ }\textbf
  {\bibinfo {volume} {434}},\ \bibinfo {pages} {484} (\bibinfo {year}
  {2005})}\BibitemShut {NoStop}%
\bibitem [{\citenamefont {Makarovski}\ \emph
  {et~al.}(2007{\natexlab{a}})\citenamefont {Makarovski}, \citenamefont
  {Zhukov}, \citenamefont {Liu},\ and\ \citenamefont
  {Finkelstein}}]{Finkelstein_PRB(75)_2007}%
  \BibitemOpen
  \bibfield  {author} {\bibinfo {author} {\bibfnamefont {A.}~\bibnamefont
  {Makarovski}}, \bibinfo {author} {\bibfnamefont {A.}~\bibnamefont {Zhukov}},
  \bibinfo {author} {\bibfnamefont {J.}~\bibnamefont {Liu}}, \ and\ \bibinfo
  {author} {\bibfnamefont {G.}~\bibnamefont {Finkelstein}},\ }\href@noop {}
  {\bibfield  {journal} {\bibinfo  {journal} {Phys. Rev. B}\ }\textbf {\bibinfo
  {volume} {75}},\ \bibinfo {pages} {241407} (\bibinfo {year}
  {2007}{\natexlab{a}})}\BibitemShut {NoStop}%
\bibitem [{\citenamefont {Makarovski}\ \emph
  {et~al.}(2007{\natexlab{b}})\citenamefont {Makarovski}, \citenamefont {Liu},\
  and\ \citenamefont {Finkelstein}}]{Makarovski_Liu_Finkelstein_PRL(99)_2007}%
  \BibitemOpen
  \bibfield  {author} {\bibinfo {author} {\bibfnamefont {A.}~\bibnamefont
  {Makarovski}}, \bibinfo {author} {\bibfnamefont {J.}~\bibnamefont {Liu}}, \
  and\ \bibinfo {author} {\bibfnamefont {G.}~\bibnamefont {Finkelstein}},\
  }\href@noop {} {\bibfield  {journal} {\bibinfo  {journal} {Phys. Rev. Lett.}\
  }\textbf {\bibinfo {volume} {99}},\ \bibinfo {pages} {066801} (\bibinfo
  {year} {2007}{\natexlab{b}})}\BibitemShut {NoStop}%
\bibitem [{\citenamefont {Ferrier}\ \emph {et~al.}(2016)\citenamefont
  {Ferrier}, \citenamefont {Arakawa}, \citenamefont {Hata}, \citenamefont
  {Fujiwara}, \citenamefont {Delagrange}, \citenamefont {Weil}, \citenamefont
  {Deblock}, \citenamefont {Sakano}, \citenamefont {Oguri},\ and\ \citenamefont
  {Kobayashi}}]{Ferrier_NAT_(12)_2016}%
  \BibitemOpen
  \bibfield  {author} {\bibinfo {author} {\bibfnamefont {M.}~\bibnamefont
  {Ferrier}}, \bibinfo {author} {\bibfnamefont {T.}~\bibnamefont {Arakawa}},
  \bibinfo {author} {\bibfnamefont {T.}~\bibnamefont {Hata}}, \bibinfo {author}
  {\bibfnamefont {R.}~\bibnamefont {Fujiwara}}, \bibinfo {author}
  {\bibfnamefont {R.}~\bibnamefont {Delagrange}}, \bibinfo {author}
  {\bibfnamefont {R.}~\bibnamefont {Weil}}, \bibinfo {author} {\bibfnamefont
  {R.}~\bibnamefont {Deblock}}, \bibinfo {author} {\bibfnamefont
  {R.}~\bibnamefont {Sakano}}, \bibinfo {author} {\bibfnamefont
  {A.}~\bibnamefont {Oguri}}, \ and\ \bibinfo {author} {\bibfnamefont
  {K.}~\bibnamefont {Kobayashi}},\ }\href@noop {} {\bibfield  {journal}
  {\bibinfo  {journal} {Nat. Phys.}\ }\textbf {\bibinfo {volume} {12}},\
  \bibinfo {pages} {230} (\bibinfo {year} {2016})}\BibitemShut {NoStop}%
\bibitem [{\citenamefont {Keller}\ \emph {et~al.}(2014)\citenamefont {Keller},
  \citenamefont {Amasha}, \citenamefont {Weymann}, \citenamefont {Moca},
  \citenamefont {Rau}, \citenamefont {Katine}, \citenamefont {Shtrikman},
  \citenamefont {Zar{\'a}nd},\ and\ \citenamefont
  {Goldhaber-Gordon}}]{Keller_Goldhaber_nat_phys(10)_2014}%
  \BibitemOpen
  \bibfield  {author} {\bibinfo {author} {\bibfnamefont {A.~J.}\ \bibnamefont
  {Keller}}, \bibinfo {author} {\bibfnamefont {S.}~\bibnamefont {Amasha}},
  \bibinfo {author} {\bibfnamefont {I.}~\bibnamefont {Weymann}}, \bibinfo
  {author} {\bibfnamefont {C.~P.}\ \bibnamefont {Moca}}, \bibinfo {author}
  {\bibfnamefont {I.~G.}\ \bibnamefont {Rau}}, \bibinfo {author} {\bibfnamefont
  {J.~A.}\ \bibnamefont {Katine}}, \bibinfo {author} {\bibfnamefont
  {H.}~\bibnamefont {Shtrikman}}, \bibinfo {author} {\bibfnamefont
  {G.}~\bibnamefont {Zar{\'a}nd}}, \ and\ \bibinfo {author} {\bibfnamefont
  {D.}~\bibnamefont {Goldhaber-Gordon}},\ }\href@noop {} {\bibfield  {journal}
  {\bibinfo  {journal} {Nature Physics}\ }\textbf {\bibinfo {volume} {10}},\
  \bibinfo {pages} {145} (\bibinfo {year} {2014})}\BibitemShut {NoStop}%
\bibitem [{\citenamefont {Bauer}\ \emph
  {et~al.}(2013{\natexlab{b}})\citenamefont {Bauer}, \citenamefont {Salomon},\
  and\ \citenamefont {Demler}}]{Bauer_Salomon_Demler_PRL_(111)_2013}%
  \BibitemOpen
  \bibfield  {author} {\bibinfo {author} {\bibfnamefont {J.}~\bibnamefont
  {Bauer}}, \bibinfo {author} {\bibfnamefont {C.}~\bibnamefont {Salomon}}, \
  and\ \bibinfo {author} {\bibfnamefont {E.}~\bibnamefont {Demler}},\
  }\href@noop {} {\bibfield  {journal} {\bibinfo  {journal} {Phys. Rev. Lett.}\
  }\textbf {\bibinfo {volume} {111}},\ \bibinfo {pages} {215304} (\bibinfo
  {year} {2013}{\natexlab{b}})}\BibitemShut {NoStop}%
\bibitem [{\citenamefont {Nishida}(2013)}]{Nishida_PRL_(111)_2013}%
  \BibitemOpen
  \bibfield  {author} {\bibinfo {author} {\bibfnamefont {Y.}~\bibnamefont
  {Nishida}},\ }\href@noop {} {\bibfield  {journal} {\bibinfo  {journal} {Phys.
  Rev. Lett.}\ }\textbf {\bibinfo {volume} {111}},\ \bibinfo {pages} {135301}
  (\bibinfo {year} {2013})}\BibitemShut {NoStop}%
\bibitem [{\citenamefont {Nishida}(2016)}]{Nishida_PRA_(93)_2016}%
  \BibitemOpen
  \bibfield  {author} {\bibinfo {author} {\bibfnamefont {Y.}~\bibnamefont
  {Nishida}},\ }\href@noop {} {\bibfield  {journal} {\bibinfo  {journal} {Phys.
  Rev. A}\ }\textbf {\bibinfo {volume} {93}},\ \bibinfo {pages} {011606(R)}
  (\bibinfo {year} {2016})}\BibitemShut {NoStop}%
\bibitem [{\citenamefont {Kuzmenko}\ \emph {et~al.}(2016)\citenamefont
  {Kuzmenko}, \citenamefont {Kuzmenko}, \citenamefont {Avishai},\ and\
  \citenamefont {Jo}}]{Kuzmenko_Avishai_2016}%
  \BibitemOpen
  \bibfield  {author} {\bibinfo {author} {\bibfnamefont {I.}~\bibnamefont
  {Kuzmenko}}, \bibinfo {author} {\bibfnamefont {T.}~\bibnamefont {Kuzmenko}},
  \bibinfo {author} {\bibfnamefont {Y.}~\bibnamefont {Avishai}}, \ and\
  \bibinfo {author} {\bibfnamefont {G.-B.}\ \bibnamefont {Jo}},\ }\href@noop {}
  {\bibfield  {journal} {\bibinfo  {journal} {Phys. Rev. B}\ }\textbf {\bibinfo
  {volume} {93}},\ \bibinfo {pages} {115143} (\bibinfo {year}
  {2016})}\BibitemShut {NoStop}%
\bibitem [{\citenamefont {Kuzmenko}\ and\ \citenamefont
  {Avishai}(2014)}]{Avishai_PRB_(89)_2014}%
  \BibitemOpen
  \bibfield  {author} {\bibinfo {author} {\bibfnamefont {I.}~\bibnamefont
  {Kuzmenko}}\ and\ \bibinfo {author} {\bibfnamefont {Y.}~\bibnamefont
  {Avishai}},\ }\href@noop {} {\bibfield  {journal} {\bibinfo  {journal} {Phys.
  Rev. B}\ }\textbf {\bibinfo {volume} {89}},\ \bibinfo {pages} {195110}
  (\bibinfo {year} {2014})}\BibitemShut {NoStop}%
\bibitem [{\citenamefont {LeHur}\ \emph {et~al.}(2004)\citenamefont {LeHur},
  \citenamefont {Simon},\ and\ \citenamefont
  {Borda}}]{Hur_Simon_Borda_PRB_(69)_2004}%
  \BibitemOpen
  \bibfield  {author} {\bibinfo {author} {\bibfnamefont {K.}~\bibnamefont
  {LeHur}}, \bibinfo {author} {\bibfnamefont {P.}~\bibnamefont {Simon}}, \ and\
  \bibinfo {author} {\bibfnamefont {L.}~\bibnamefont {Borda}},\ }\href@noop {}
  {\bibfield  {journal} {\bibinfo  {journal} {Phys. Rev. B}\ }\textbf {\bibinfo
  {volume} {69}},\ \bibinfo {pages} {045326} (\bibinfo {year}
  {2004})}\BibitemShut {NoStop}%
\bibitem [{\citenamefont {Lopez}\ \emph {et~al.}(2005)\citenamefont {Lopez},
  \citenamefont {Sanchez}, \citenamefont {Lee}, \citenamefont {Choi},
  \citenamefont {Simon},\ and\ \citenamefont
  {LeHur}}]{Lopez_Hur_PRB_(71)_2005}%
  \BibitemOpen
  \bibfield  {author} {\bibinfo {author} {\bibfnamefont {R.}~\bibnamefont
  {Lopez}}, \bibinfo {author} {\bibfnamefont {D.}~\bibnamefont {Sanchez}},
  \bibinfo {author} {\bibfnamefont {M.}~\bibnamefont {Lee}}, \bibinfo {author}
  {\bibfnamefont {M.-S.}\ \bibnamefont {Choi}}, \bibinfo {author}
  {\bibfnamefont {P.}~\bibnamefont {Simon}}, \ and\ \bibinfo {author}
  {\bibfnamefont {K.}~\bibnamefont {LeHur}},\ }\href@noop {} {\bibfield
  {journal} {\bibinfo  {journal} {Phys. Rev. B}\ }\textbf {\bibinfo {volume}
  {71}},\ \bibinfo {pages} {115312} (\bibinfo {year} {2005})}\BibitemShut
  {NoStop}%
\bibitem [{\citenamefont {Schmid}\ \emph {et~al.}(2015)\citenamefont {Schmid},
  \citenamefont {Smirnov}, \citenamefont {Marganska}, \citenamefont
  {Dirnaichner}, \citenamefont {Stiller}, \citenamefont {Grifoni},
  \citenamefont {H\"uttel},\ and\ \citenamefont
  {Strunk}}]{Schmid_Grifoni_PRB_(91)_2015}%
  \BibitemOpen
  \bibfield  {author} {\bibinfo {author} {\bibfnamefont {D.~R.}\ \bibnamefont
  {Schmid}}, \bibinfo {author} {\bibfnamefont {S.}~\bibnamefont {Smirnov}},
  \bibinfo {author} {\bibfnamefont {M.}~\bibnamefont {Marganska}}, \bibinfo
  {author} {\bibfnamefont {A.}~\bibnamefont {Dirnaichner}}, \bibinfo {author}
  {\bibfnamefont {P.~L.}\ \bibnamefont {Stiller}}, \bibinfo {author}
  {\bibfnamefont {M.}~\bibnamefont {Grifoni}}, \bibinfo {author} {\bibfnamefont
  {A.~K.}\ \bibnamefont {H\"uttel}}, \ and\ \bibinfo {author} {\bibfnamefont
  {C.}~\bibnamefont {Strunk}},\ }\href@noop {} {\bibfield  {journal} {\bibinfo
  {journal} {Phys. Rev. B}\ }\textbf {\bibinfo {volume} {91}},\ \bibinfo
  {pages} {155435} (\bibinfo {year} {2015})}\BibitemShut {NoStop}%
\bibitem [{\citenamefont {Sakano}\ and\ \citenamefont
  {Kawakami}(2006)}]{Sakano_JMMM_(310)_2006}%
  \BibitemOpen
  \bibfield  {author} {\bibinfo {author} {\bibfnamefont {R.}~\bibnamefont
  {Sakano}}\ and\ \bibinfo {author} {\bibfnamefont {N.}~\bibnamefont
  {Kawakami}},\ }\href@noop {} {\bibfield  {journal} {\bibinfo  {journal} {J.
  Mag. Mag. Matt.}\ }\textbf {\bibinfo {volume} {310}},\ \bibinfo {pages}
  {1136} (\bibinfo {year} {2006})}\BibitemShut {NoStop}%
\bibitem [{\citenamefont {Sakano}\ \emph {et~al.}(2007)\citenamefont {Sakano},
  \citenamefont {Kita},\ and\ \citenamefont
  {Kawakami}}]{Sakano_JPSJ_(76)_2007}%
  \BibitemOpen
  \bibfield  {author} {\bibinfo {author} {\bibfnamefont {R.}~\bibnamefont
  {Sakano}}, \bibinfo {author} {\bibfnamefont {T.}~\bibnamefont {Kita}}, \ and\
  \bibinfo {author} {\bibfnamefont {N.}~\bibnamefont {Kawakami}},\ }\href@noop
  {} {\bibfield  {journal} {\bibinfo  {journal} {J. Phys. Soc. Jpn}\ }\textbf
  {\bibinfo {volume} {76}},\ \bibinfo {pages} {074709} (\bibinfo {year}
  {2007})}\BibitemShut {NoStop}%
\bibitem [{\citenamefont {Roura-Bas}\ \emph {et~al.}(2012)\citenamefont
  {Roura-Bas}, \citenamefont {Tosi}, \citenamefont {Aligia},\ and\
  \citenamefont {Cornaglia}}]{Tosi_PRB_(86)_2012}%
  \BibitemOpen
  \bibfield  {author} {\bibinfo {author} {\bibfnamefont {P.}~\bibnamefont
  {Roura-Bas}}, \bibinfo {author} {\bibfnamefont {L.}~\bibnamefont {Tosi}},
  \bibinfo {author} {\bibfnamefont {A.~A.}\ \bibnamefont {Aligia}}, \ and\
  \bibinfo {author} {\bibfnamefont {P.~S.}\ \bibnamefont {Cornaglia}},\
  }\href@noop {} {\bibfield  {journal} {\bibinfo  {journal} {Phys. Rev. B}\
  }\textbf {\bibinfo {volume} {86}},\ \bibinfo {pages} {165106} (\bibinfo
  {year} {2012})}\BibitemShut {NoStop}%
{\color{black} \bibitem [{\citenamefont {Azema}\ \emph {et~al.}(2012)\citenamefont {Azema},
  \citenamefont {Dar{\'e}}, \citenamefont {Sch{\"a}fer},\ and\ \citenamefont
  {Lombardo}}]{Azema_PRB(86)_2012}%
  \BibitemOpen
  \bibfield  {author} {\bibinfo {author} {\bibfnamefont {J.}~\bibnamefont
  {Azema}}, \bibinfo {author} {\bibfnamefont {A.-M.}\ \bibnamefont {Dar{\'e}}},
  \bibinfo {author} {\bibfnamefont {S.}~\bibnamefont {Sch{\"a}fer}}, \ and\
  \bibinfo {author} {\bibfnamefont {P.}~\bibnamefont {Lombardo}},\ }\href@noop
  {} {\bibfield  {journal} {\bibinfo  {journal} {Phys. Rev. B}\ }\textbf
  {\bibinfo {volume} {86}},\ \bibinfo {pages} {075303} (\bibinfo {year}
  {2012})}}\BibitemShut {NoStop}%
\bibitem [{\citenamefont {Dorda}\ \emph {et~al.}(2016)\citenamefont {Dorda},
  \citenamefont {Ganahl}, \citenamefont {Andergassen}, \citenamefont {von~der
  Linden},\ and\ \citenamefont {Arrigoni}}]{Andergassen_PRB_(94)_2016}%
  \BibitemOpen
  \bibfield  {author} {\bibinfo {author} {\bibfnamefont {A.}~\bibnamefont
  {Dorda}}, \bibinfo {author} {\bibfnamefont {M.}~\bibnamefont {Ganahl}},
  \bibinfo {author} {\bibfnamefont {S.}~\bibnamefont {Andergassen}}, \bibinfo
  {author} {\bibfnamefont {W.}~\bibnamefont {von~der Linden}}, \ and\ \bibinfo
  {author} {\bibfnamefont {E.}~\bibnamefont {Arrigoni}},\ }\href@noop {}
  {\bibfield  {journal} {\bibinfo  {journal} {Phys. Rev. B}\ }\textbf {\bibinfo
  {volume} {94}},\ \bibinfo {pages} {245125} (\bibinfo {year}
  {2016})}\BibitemShut {NoStop}%
\bibitem [{\citenamefont {Schrieffer}\ and\ \citenamefont
  {Wolf}(1966)}]{Schrieffer_Wolf_PR(149)_1966}%
  \BibitemOpen
  \bibfield  {author} {\bibinfo {author} {\bibfnamefont {J.~R.}\ \bibnamefont
  {Schrieffer}}\ and\ \bibinfo {author} {\bibfnamefont {P.}~\bibnamefont
  {Wolf}},\ }\href@noop {} {\bibfield  {journal} {\bibinfo  {journal} {Phys.
  Rev.}\ }\textbf {\bibinfo {volume} {149}},\ \bibinfo {pages} {491} (\bibinfo
  {year} {1966})}\BibitemShut {NoStop}%
\bibitem [{\citenamefont {Glazman}\ and\ \citenamefont
  {Raikh}(1988)}]{Glazman_Raikh_JETP_(27)_1988}%
  \BibitemOpen
  \bibfield  {author} {\bibinfo {author} {\bibfnamefont {L.~I.}\ \bibnamefont
  {Glazman}}\ and\ \bibinfo {author} {\bibfnamefont {M.~E.}\ \bibnamefont
  {Raikh}},\ }\href@noop {} {\bibfield  {journal} {\bibinfo  {journal} {J. Exp.
  Theor. Phys.}\ }\textbf {\bibinfo {volume} {27}},\ \bibinfo {pages} {452}
  (\bibinfo {year} {1988})}\BibitemShut {NoStop}%
\bibitem [{\citenamefont {Ng}\ and\ \citenamefont
  {Lee}(1988)}]{Ng_Lee_PRL_(61)_1988}%
  \BibitemOpen
  \bibfield  {author} {\bibinfo {author} {\bibfnamefont {T.~K.}\ \bibnamefont
  {Ng}}\ and\ \bibinfo {author} {\bibfnamefont {P.~A.}\ \bibnamefont {Lee}},\
  }\href@noop {} {\bibfield  {journal} {\bibinfo  {journal} {Phys. Rev. Lett.}\
  }\textbf {\bibinfo {volume} {61}},\ \bibinfo {pages} {1768} (\bibinfo {year}
  {1988})}\BibitemShut {NoStop}%
\bibitem [{\citenamefont {Pustilnik}\ and\ \citenamefont
  {Glazman}(2004)}]{GP_Review_2005}%
  \BibitemOpen
  \bibfield  {author} {\bibinfo {author} {\bibfnamefont {M.}~\bibnamefont
  {Pustilnik}}\ and\ \bibinfo {author} {\bibfnamefont {L.}~\bibnamefont
  {Glazman}},\ }\href@noop {} {\bibfield  {journal} {\bibinfo  {journal} {J.
  Phys.: Condens. Matter}\ }\textbf {\bibinfo {volume} {16}},\ \bibinfo {pages}
  {R513} (\bibinfo {year} {2004})}\BibitemShut {NoStop}%
\bibitem [{Note1()}]{Note1}%
  \BibitemOpen
  \bibinfo {note} {The weak coupling Hamiltonian (\ref {HamKon}) sets the Kondo
  temperature $T_K \sim D\protect \qopname \relax o{exp}\left [-1/(N\nu
  J)\right ]$. Here $D$ is a bandwidth of conduction electrons
  band.}\BibitemShut {Stop}%
\bibitem [{\citenamefont {Karki}\ and\ \citenamefont
  {Kiselev}()}]{Karki_MK_Supplement}%
  \BibitemOpen
  \bibfield  {author} {\bibinfo {author} {\bibfnamefont {D.}~\bibnamefont
  {Karki}}\ and\ \bibinfo {author} {\bibfnamefont {M.~N.}\ \bibnamefont
  {Kiselev}},\ }\href@noop {} {\bibinfo  {journal} {Supplemental Materials}\
  }\BibitemShut {NoStop}%
\bibitem [{Note2()}]{Note2}%
  \BibitemOpen
\bibfield  {journal} {  }\bibinfo {note} {We omitted the six-fermion term \cite
  { Mora_Clerk_Hur_PRB(80)_2009} in (\ref {HamFL}) since it produces
  perturbative corrections to the current beyond the accuracy of our
  theory.}\BibitemShut {Stop}%
\bibitem [{\citenamefont {Keldysh}(1965)}]{Keldysh_SP_JETP(20)_1965}%
  \BibitemOpen
  \bibfield  {author} {\bibinfo {author} {\bibfnamefont {L.~V.}\ \bibnamefont
  {Keldysh}},\ }\href@noop {} {\bibfield  {journal} {\bibinfo  {journal} {Sov.
  Phys. JETP}\ }\textbf {\bibinfo {volume} {20}},\ \bibinfo {pages} {1018}
  (\bibinfo {year} {1965})}\BibitemShut {NoStop}%
\bibitem [{Note3()}]{Note3}%
  \BibitemOpen
  \bibinfo {note} {Approaches based on the self-energies $\Sigma $ and $T$-
  matrix are equivalent since $\Sigma (\epsilon )=T(\epsilon )(1+{\protect \cal
  G}_0(\epsilon )\cdot T(\epsilon ))^{-1}$, see \cite {Karki_MK_Supplement} for
  details.}\BibitemShut {Stop}%
{\color{black} \bibitem [{\citenamefont {Ferrier}\ \emph {et~al.}(2017)\citenamefont
  {Ferrier}, \citenamefont {Arakawa}, \citenamefont {Hata}, \citenamefont
  {Fujiwara}, \citenamefont {Delagrange}, \citenamefont {Deblock},
  \citenamefont {Teratani}, \citenamefont {Sakano}, \citenamefont {Oguri},\
  and\ \citenamefont {Kobayashi}}]{Ferrier_PRL_2017}%
  \BibitemOpen
  \bibfield  {author} {\bibinfo {author} {\bibfnamefont {M.}~\bibnamefont
  {Ferrier}}, \bibinfo {author} {\bibfnamefont {T.}~\bibnamefont {Arakawa}},
  \bibinfo {author} {\bibfnamefont {T.}~\bibnamefont {Hata}}, \bibinfo {author}
  {\bibfnamefont {R.}~\bibnamefont {Fujiwara}}, \bibinfo {author}
  {\bibfnamefont {R.}~\bibnamefont {Delagrange}}, \bibinfo {author}
  {\bibfnamefont {R.}~\bibnamefont {Deblock}}, \bibinfo {author} {\bibfnamefont
  {Y.}~\bibnamefont {Teratani}}, \bibinfo {author} {\bibfnamefont
  {R.}~\bibnamefont {Sakano}}, \bibinfo {author} {\bibfnamefont
  {A.}~\bibnamefont {Oguri}}, \ and\ \bibinfo {author} {\bibfnamefont
  {K.}~\bibnamefont {Kobayashi}},\ }\href@noop {} {\bibfield  {journal}
  {\bibinfo  {journal} {Phys. Rev. Lett.}\ }\textbf {\bibinfo {volume} {118}},\
  \bibinfo {pages} {196803} (\bibinfo {year} {2017})}}\BibitemShut {NoStop}%
\bibitem [{Note4()}]{Note4}%
  \BibitemOpen
  \bibinfo {note} {It is instructive to re-write conductance in (\ref
  {elecondu}) as $G(T)$$=$$G(0)$$($$1$$-$$c_T$$(\pi T/T_K)^2)$,
  $G$$($$0)$$=$$G_0$$\protect \qopname \relax o{sin}^2x$ and the FL constant
  $c_T$$=$$($$N$$+$$1$$)$$/$$($$3$$($$N$$-$$1$$)$$)$$($$2$$\protect \qopname
  \relax o{cos}$$ 2x$$)$$/($$\protect \qopname \relax o{cos}$$ 2x$$-$$1$$)$.
  Potential scattering results in the replacement $x$$=$$\delta _0$$\to
  $$\delta _0$$+$$\delta _P$ and renormalizes both $G$$($$0)$ and $c_T$ \cite
  {GP_Review_2005}.}\BibitemShut {Stop}%
\bibitem [{Note5()}]{Note5}%
  \BibitemOpen
  \bibinfo {note} {This relation is also know as the Mott-Cutler formula \cite
  {Costi_Zlatic_PRB(81)_2010}.}\BibitemShut {Stop}%
\bibitem [{Note6()}]{Note6}%
  \BibitemOpen
  \bibinfo {note} {The offset can be easily understood on trivial example of
  SU(4) $m$$=$$1$ and $\delta _P$$=$$0$. In that case the inelastic
  contribution to the current vanish and the non-linear effect is $\propto $
  $(\Delta T)^2$. Therefore, the offset is linear in $\Delta T$ and can be used
  as a measure of the temperature drop.}\BibitemShut {Stop}%
\bibitem [{\citenamefont {Luttinger}(1964)}]{Luttinger_PR_(135)_1964}%
  \BibitemOpen
  \bibfield  {author} {\bibinfo {author} {\bibfnamefont {J.~M.}\ \bibnamefont
  {Luttinger}},\ }\href@noop {} {\bibfield  {journal} {\bibinfo  {journal}
  {Phys. Rev.}\ }\textbf {\bibinfo {volume} {135}},\ \bibinfo {pages} {016501}
  (\bibinfo {year} {1964})}\BibitemShut {NoStop}%
\bibitem [{\citenamefont {Shastry}(2009)}]{Shastry_RPP_(72)_2009}%
  \BibitemOpen
  \bibfield  {author} {\bibinfo {author} {\bibfnamefont {B.~S.}\ \bibnamefont
  {Shastry}},\ }\href@noop {} {\bibfield  {journal} {\bibinfo  {journal} {Rep.
  Prog. Phys.}\ }\textbf {\bibinfo {volume} {72}},\ \bibinfo {pages} {016501}
  (\bibinfo {year} {2009})}\BibitemShut {NoStop}%
\bibitem [{\citenamefont {Eich}\ \emph {et~al.}(2014)\citenamefont {Eich},
  \citenamefont {Principi}, \citenamefont {DiVentra},\ and\ \citenamefont
  {Vignale}}]{Eich_PRB_(90)_2014}%
  \BibitemOpen
  \bibfield  {author} {\bibinfo {author} {\bibfnamefont {F.~G.}\ \bibnamefont
  {Eich}}, \bibinfo {author} {\bibfnamefont {A.}~\bibnamefont {Principi}},
  \bibinfo {author} {\bibfnamefont {M.}~\bibnamefont {DiVentra}}, \ and\
  \bibinfo {author} {\bibfnamefont {G.}~\bibnamefont {Vignale}},\ }\href@noop
  {} {\bibfield  {journal} {\bibinfo  {journal} {Phys. Rev. B}\ }\textbf
  {\bibinfo {volume} {90}},\ \bibinfo {pages} {115116} (\bibinfo {year}
  {2014})}\BibitemShut {NoStop}%
\bibitem [{\citenamefont {Karki}\ and\ \citenamefont
  {Kiselev}(2017)}]{Karki_MK_TBP}%
  \BibitemOpen
  \bibfield  {author} {\bibinfo {author} {\bibfnamefont {D.}~\bibnamefont
  {Karki}}\ and\ \bibinfo {author} {\bibfnamefont {M.~N.}\ \bibnamefont
  {Kiselev}},\ }\href@noop {} {\bibfield  {journal} {\bibinfo  {journal} {in
  preparation}\ } (\bibinfo {year} {2017})}\BibitemShut {NoStop}%
\bibitem [{\citenamefont {Costi}\ and\ \citenamefont
  {Zlati{\'c}}(2010)}]{Costi_Zlatic_PRB(81)_2010}%
  \BibitemOpen
  \bibfield  {author} {\bibinfo {author} {\bibfnamefont {T.~A.}\ \bibnamefont
  {Costi}}\ and\ \bibinfo {author} {\bibfnamefont {V.}~\bibnamefont
  {Zlati{\'c}}},\ }\href@noop {} {\bibfield  {journal} {\bibinfo  {journal}
  {Phys. Rev. B}\ }\textbf {\bibinfo {volume} {81}},\ \bibinfo {pages} {235127}
  (\bibinfo {year} {2010})}\BibitemShut {NoStop}%
\bibitem [{Note7()}]{Note7}%
  \BibitemOpen
  \bibinfo {note} {The Lorenz number $L_0(T)$ for given $T$ connects the
  thermal $K_e(T)$ and electrical $G(T)$ conductances:
  $L_0$$($$T$$)$$=$$K_e$$($$T$$)$$/$$T$$G$$(T)$ \cite {Karki_MK_Supplement}.
  The figure of merit (neglecting the phonon contribution) is defined by
  $ZT$$=$$S^2(T)$$/$$L_0(T)$. The highest $ZT$ is achieved in the PHN regime at
  $\delta _P$$=$$0$ and symmetric dot-leads coupling.}\BibitemShut {Stop}%
{\color{black} 
\bibitem [{\citenamefont {Sierra}\ \emph {et~al.}(2017)\citenamefont {Sierra},
  \citenamefont {Lopez},\ and\ \citenamefont {Sanchez}}]{Lopez_arXiv_2017}%
  \BibitemOpen
  \bibfield  {author} {\bibinfo {author} {\bibfnamefont {M.~A.}\ \bibnamefont
  {Sierra}}, \bibinfo {author} {\bibfnamefont {R.}~\bibnamefont {Lopez}}, \
  and\ \bibinfo {author} {\bibfnamefont {D.}~\bibnamefont {Sanchez}},\
  }\href@noop {} {\bibfield  {journal} {\bibinfo  {journal} {Phys. Rev. B}\
  }\textbf {\bibinfo {volume} {96}},\ \bibinfo {pages} {085416} (\bibinfo
  {year} {2017})}}\BibitemShut {NoStop}%

\end{thebibliography}

%

\begin{widetext}
\setcounter{equation}{0}
\setcounter{figure}{0}
\setcounter{table}{0}
\setcounter{page}{1}
\makeatletter
\renewcommand{\theequation}{S\arabic{equation}}
\renewcommand{\thefigure}{S\arabic{figure}}
\renewcommand{\bibnumfmt}[1]{[S#1]}
\renewcommand{\citenumfont}[1]{S#1}
\section*{Supplemental Materials}\label{i}
This Supplemental Materials contains additional information about the charge current beyond the linear response theory and connections between the full fledged calculations performed using Kedlysh 
out-of-equilibrium approach and the results derived by means of the transport integrals method.
\section{Electric current beyond the linear response}
\subsection{Coupling asymmetry}
If the quantum impurity is coupled to the leads with arbitrary coupling, the new variables ($a$ and $b$) entering the FL Hamiltonian (\ref{HamFL}) are defined by Glazman-Raikh rotation \cite{Glazman_Raikh_JETP_(27)_1988,Ng_Lee_PRL_(61)_1988,GP_Review_2005} as follows:
\begin{equation}
 \left(%
\begin{array}{c}
  b_{k r} \\
  a_{k r} \\
\end{array}%
\right)= \left(%
\begin{array}{cc}
  \cos\theta & \sin\theta \\
  \sin\theta & -\cos\theta \\
\end{array}%
\right)\left(%
\begin{array}{c}
  c_{Lk r} \\
  c_{Rk r} \\
\end{array}%
\right),
\end{equation}
where $\tan$$\theta$$=$$|t_R/t_L|$ is given by the ratio of tunnel lead-dot matrix elements \cite{GP_Review_2005}
of the Hamiltonian (\ref{HamAnd}). The symmetric coupling corresponds to $\theta$$=$$\pi/4$. We introduce the parameter $\mathcal{C}$$=$$\cos2\theta$$=$$(\Gamma_L-\Gamma_R)$$/$$(\Gamma_L+\Gamma_R)$ to characterize the asymmetry of the dot-lead coupling; $\Gamma_\alpha$$=$$\pi$$ \nu$$ N $$|t_\alpha|^2$ is intrinsic total local level width associated with the tunneling from/to the reservoirs (we assume that tunnel matrix elements are the same for all orbitals/flavours).
\subsection{Phase shift}
\noindent The phase shift expression in the presence of the finite voltage bias $e\Delta V$, finite temperature drop across the impurity $\Delta T$ is given by the equation \cite{Mora_Clerk_Hur_PRB(80)_2009}:
\begin{equation}
\label{phase_shift}
\delta (\varepsilon)=\delta_0+\alpha_1\varepsilon+\alpha_2\varepsilon^2-\frac{N-1}{4}\phi_2\cdot {\cal A},
\end{equation}
\noindent where
\begin{equation}\label{addition}
{\cal A}=\frac{1}{6}\left[(\pi T_L)^2(1+\mathcal{C})+(\pi T_R)^2(1-\mathcal{C})+\frac{3}{2}(1-\mathcal{C}^2)(e\Delta V)^2\right].
\end{equation}
\noindent The FL identities $\alpha_1=(N-1) \phi_1$ and  $\alpha_2=1/4 (N-1)\phi_2$ follow from the Kondo floating paradigm \cite{Nozieres,Mora_Clerk_Hur_PRB(80)_2009,Mora_PRB_(80)_2009}.
The exact relation between $\alpha_1$ and $\alpha_2$ is given by the Bethe-Ansatz solution  \cite{Mora_Clerk_Hur_PRB(80)_2009,Mora_PRB_(80)_2009}:
\begin{equation}
\frac{\alpha_2}{\alpha^2_1}=\frac{N-2}{N-1}\frac{\Gamma(1/N)\tan(\pi/N)}{\sqrt{\pi}\Gamma\left(\frac{1}{2}+\frac{1}{N}\right)}\cot\delta_0.
\end{equation}
Here $\Gamma(x)$ is the Euler's gamma-function.
\subsection{Elastic current}
\noindent The elastic contribution to the charge current is performed by averaging the current with 
$H_\alpha$ of (\ref{HamFL}) which is equivalent to use of the Landauer-B\"uttiker formula \cite{Blanter_Nazarov} containing  the energy dependent transmission coefficient ${\cal T}(\varepsilon)$ 
computed with the scattering phase (\ref{phase_shift}, \ref{addition}): 
\begin{equation}\label{LB_formula}
I_{el}=\frac{N e}{h}\int d\varepsilon {\cal T}(\varepsilon) \Delta f(\varepsilon), \quad 
{\cal T}(\varepsilon)=\left(1-\mathcal{C}^2\right)\sin^2[\delta(\varepsilon)].
\end{equation}
\noindent Computing integrals with Fermi distribution function we obtain the elastic current:
\begin{eqnarray}\label{current_elastic}
I_{el}=\frac{Ne}{h}\left(1-\mathcal{C}^2\right)\left[(\mathcal{T}_0-\sin2\delta_0\alpha_2 \mathcal{A})\mathcal{J}_0-\sin2\delta_0\alpha_1 \mathcal{J}_1+(\cos2\delta_0\alpha^2_1+\sin2\delta_0\alpha_2)\mathcal{J}_2\right],
\end{eqnarray}
\noindent where we use short-hand notations:
\begin{eqnarray}\label{dk2}
\mathcal{J}_0&=&e\Delta V,\quad \mathcal{J}_1=\frac{1}{6}\left[(\pi T_L)^2-(\pi T_R)^2-3 (e\Delta V)^2 \mathcal{C}\right], \nonumber\\
\mathcal{J}_2&=&\frac{e\Delta V}{6}\left[(\pi T_L)^2(1-\mathcal{C)}+(\pi T_R)^2(1+\mathcal{C)}+\frac{1}{2}(e\Delta V)^2(1+3\mathcal{C}^2)\right].
\end{eqnarray}
\subsection{Inelastic current}

\begin{figure}[b]
\begin{center}
 \includegraphics[height=50mm]{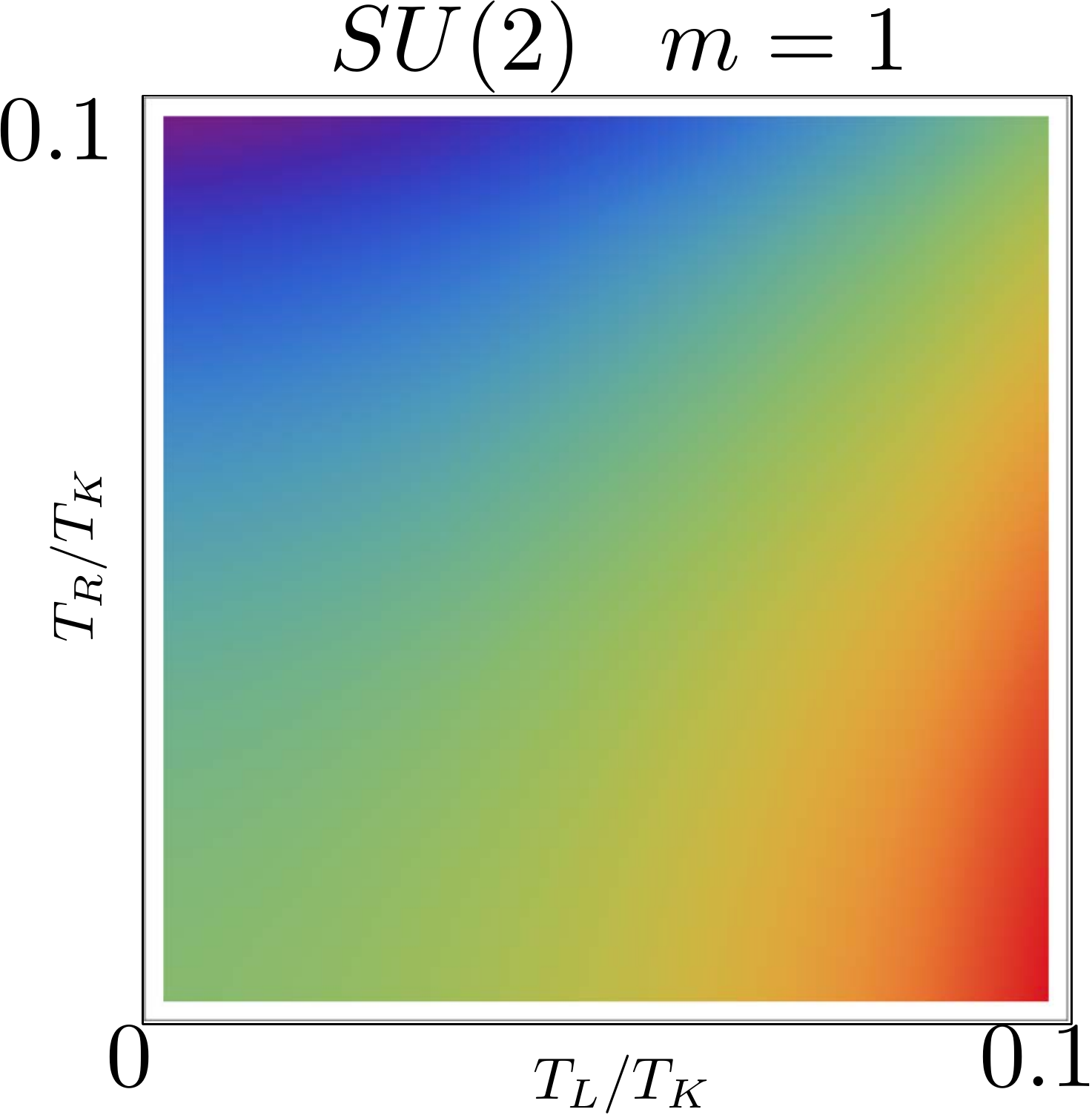}
 \includegraphics[height=50mm]{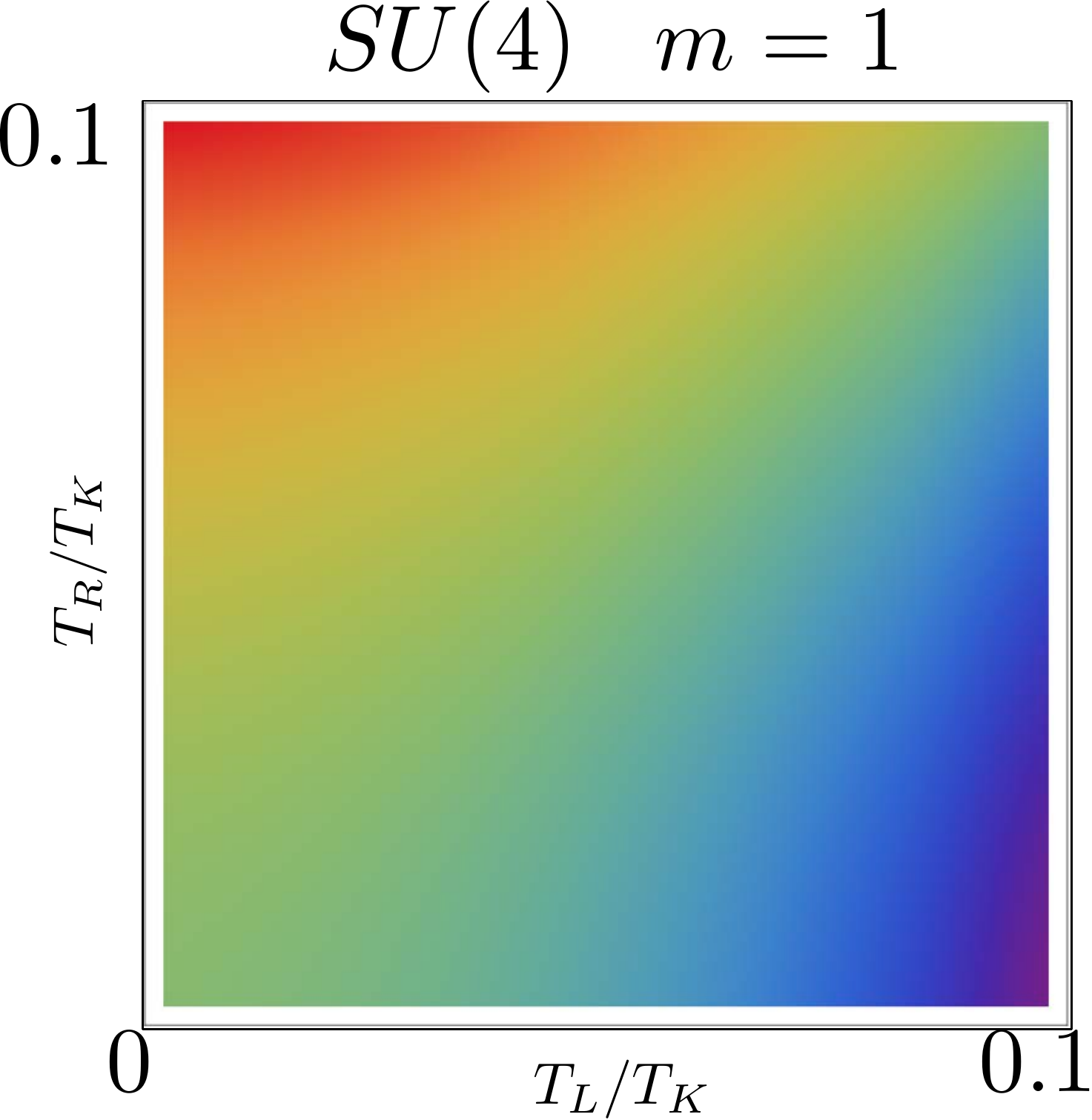}
 \includegraphics[height=50mm]{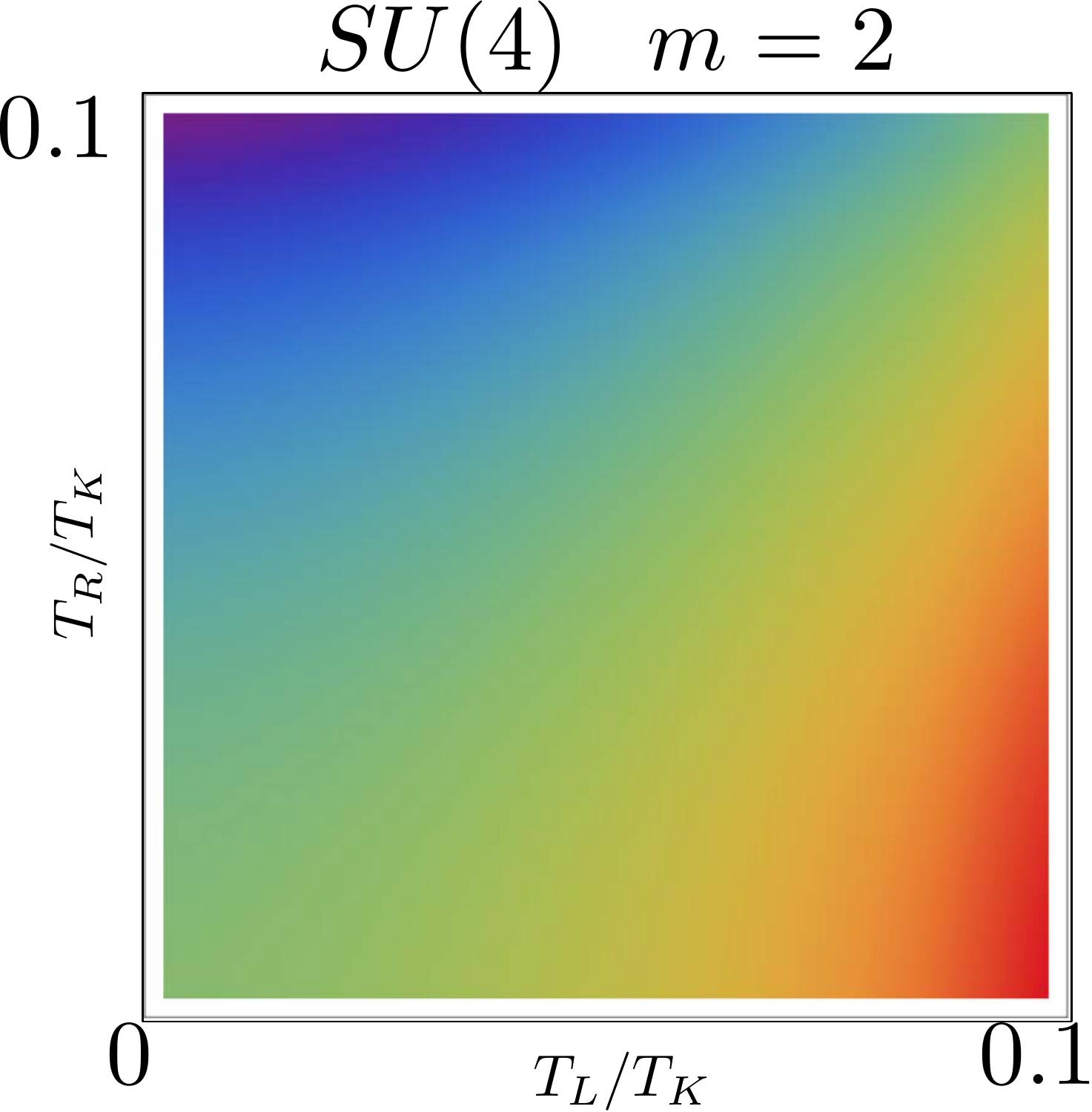}\;\;\;\;\;\;
 \includegraphics[height=50mm]{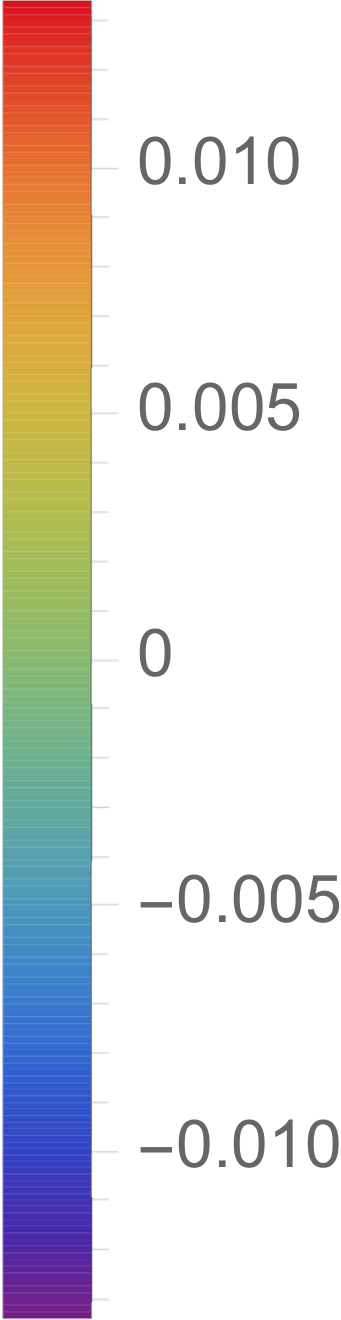}
\caption{(Color online) Density plot showing the thermo-voltage $e\Delta V/T_K$ obtained at the zero-current conditions as a function of the temperatures of L/R contacts. Left panel: SU(2) PH-symmetric Kondo regime $m=1$; central panel: SU(4) PH-non-symmetric Kondo regime $m=1$; right panel: SU(4) PH-symmetric Kondo regime $m=2$; for all plots: $\delta_P$$=$$0.3$, $\gamma$$T_K= $$0.001$ and $\mathcal{C}=0$.}
\label{f.s1}
\end{center} 
\end{figure}

For computing the inelastic contribution to the current we use the general equation for the 
self-energies
\begin{eqnarray}
\Sigma^{\eta_1, \eta_2}(t)=\left(\frac{\phi_1}{\pi\nu^2}\right)^2\sum_{k_1, k_2, k_3} G^{\eta_1, \eta_2}_{bb}(k_1, t) G^{\eta_2, \eta_1}_{bb}(k_2, -t) G^{\eta_1, \eta_2}_{bb}(k_3, t)
\end{eqnarray}
expressed in terms of the fermionic Green's functions (which replace (\ref{dec11})):
\begin{eqnarray}\label{dec11s}
G^{+-}_{bb}(t)&=&-\frac{\pi\nu}{2}\left[\frac{T_L (1+\mathcal{C}) e^{-i\mu_L
t}}{\sinh(\pi T_L t)}+\frac{T_R(1-\mathcal{C}) e^{-i\mu_R t}}{\sinh(\pi
T_R t)}\right],\\
G_{ab/ba}(t)&=&-\frac{\pi\nu}{2}\sin2\theta\left[\frac{T_L e^{-i\mu_L
t}}{\sinh(\pi T_L t)}-\frac{T_R e^{-i\mu_R t}}{\sinh(\pi T_R t)}\right],
\label{dec13}
\end{eqnarray}
where
$\mu_L=\frac{e\Delta V}{2} (1-\mathcal{C})$ and  $\mu_R=-\frac{e\Delta V}{2} (1+\mathcal{C})$. We also take
into account the coefficient $\sin 2\theta$ in front of the definition of the total current in Eq. \ref{aaa} and $\sin^2 2\theta$ in front of the inelastic current in Eq. \ref{lc}.

The inelastic current for symmetrical dot-lead coupling is given by the relation:
\begin{equation}\label{current_inelastic}
\delta I_{in}=\frac{N(N-1)e\pi}{2h}\left(\frac{\phi_1}{\pi\nu^2 }\right)^2 4\cos2\delta_0 \left(\frac{\pi \nu}{2}\right)^4\times 
\left\{\left[\mathcal{L}(T_L,T_R,z)-\mathcal{L}(T_L,T_R,0)\right]-\left[\mathcal{L}(T_R,T_L,-z)-\mathcal{L}(T_R,T_L,0)\right]\right\},
\end{equation}
\noindent where
\begin{equation}\label{L_1}
\mathcal{L}(x, y, z)=\int^{+\infty-i\gamma}_{-\infty-i\gamma}\left[\frac{x^4}{\sinh^4(\pi x t)}+2x^3y\frac{e^{izt}}{\sinh^3(\pi x t) \sinh(\pi y t)}+x^2y^2\frac{e^{2izt}}{\sinh^2(\pi x t) \sinh^2(\pi y t)}\right]dt
\end{equation}

\noindent Here we denoted $z$$=$$e\Delta V$ and introduced the point splitting parameter $\gamma$ \cite{Affleck_Lud_PRB(48)_1993,Mora_Clerk_Hur_PRB(80)_2009} to regularize the integrals
(\ref{L_1}) divergent at $t$$=$$0$. The parameter $\gamma$
is chosen to satisfy the conditions $\gamma$$ e \Delta$$ V$$ \ll 1$, $\gamma$$T$$\ll 1$  and 
$T_K/D $$\ll$$ \gamma$$ T_K$$ \ll 1$ \cite{Mora_Clerk_Hur_PRB(80)_2009} ($D$ is a bandwidth
of conduction band, $T \ll T_K$, $T_K/D \propto \sqrt{(\Gamma_L+\Gamma_R)/D}\exp[- c\cdot U/(\Gamma_L+\Gamma_R)]$, $c\sim 1$).

We show on Fig. \ref{f.s1} the thermo-voltage $\Delta V$ as a function of two temperatures $T_L$ and $T_R$ of 
the left-right leads for three important cases discussed in the paper: i) $m=1$ SU(2); ii) $m=1$ SU(4) 
and iii) $m=2$ SU(4). One can see similarity of the plots i) and iii) describing the broken
by potential scattering particle-hole symmetric regimes. The density plot visualises the non-linearity
of the thermo-voltage at low compared to $T_K$ temperatures.

\section{Transport Integrals}
We illustrate the application of the textbook \cite{Zlatic, Costi_Zlatic_PRB(81)_2010} method of transport integrals to the thermoelectric transport through the SU(N) quantum impurity 
assuming the symmetric dot-leads coupling for simplicity. 
The charge and the heat currents in the linear response theory are connected by equations:
\begin{equation}
 \left(%
\begin{array}{c}
  I_{charge} \\
  I_{heat} \\
\end{array}%
\right)= \left(%
\begin{array}{cc}
  L_{11} & L_{12} \\
  L_{21} & L_{22} \\
\end{array}%
\right)\left(%
\begin{array}{c}
  \Delta V \\
  \Delta T \\
\end{array}%
\right).
\end{equation}
Differential conductance $G(T)$ and differential thermopower $S(T)$ are defined as follows:
\begin{align}
G(T)&=\lim_{\Delta V\to 0}\left.\frac{I_{charge}}{\Delta V}\right|_{\Delta T=0}= L_{11},\\
S(T)&=-\lim_{\Delta T\to 0}\left.\frac{\Delta V}{\Delta T}\right|_{I_{charge}=0}= L_{12}/L_{11}.
\end{align}

The coefficients $L_{ij}$ are expressed in terms of the transport integrals (see e.g.\cite{Costi_Zlatic_PRB(81)_2010} for details of the derivation):
\begin{equation}
{\cal I}_n(T)=\frac{1}{h}\sum_\sigma\int d\varepsilon\cdot \varepsilon^n\left(-\frac{\partial f}{\partial \varepsilon}\right)\cdot
{\rm Im}\left[-\pi \nu T_\sigma(\varepsilon)\right].
\end{equation}
Here $L_{11}=e^2 {\cal I}_0$, $L_{12} =-e {\cal I}_1/(2 T)$ and $T_\sigma$ is a diagonal part of a single-particle T-matrix defined by the Dyson equation \cite{GP_Review_2005}:
\begin{equation}
{\cal G}_{\sigma k}(\varepsilon)={\cal G}^0_{\sigma k}(\varepsilon)+{\cal G}^0_{\sigma k}(\varepsilon) T_{\sigma}(\varepsilon){\cal G}^0_{\sigma k}(\varepsilon),
\end{equation} 
where ${\cal G}^0_{\sigma k}(\varepsilon)$  and ${\cal G}_{\sigma k}(\varepsilon)$ are bare and full electron Green's functions.

The full T-matrix consists of the elastic part 
\begin{eqnarray}
T^{el}_{\sigma}(\varepsilon)=-\frac{i}{2\pi \nu}\left(1-e^{2i\delta_{\sigma}(\varepsilon)}\right),
\end{eqnarray}
and the inelastic part  
\begin{eqnarray}T^{in}_{\sigma}(\varepsilon)=-\frac{i}{2\pi\nu}(N-1)e^{2i\delta_0}\phi^2_1\left[\varepsilon^2+(\pi T)^2\right].
\end{eqnarray}
Here we used the FL Hamiltonian (\ref{HamFL}) describing the SU(N) Kondo model at the strong coupling fixed point. We compute the transport integrals by Taylor-expanding the T-matrix
\begin{equation}\label{pse}
-\pi \nu {\rm Im} T_{\sigma}(\varepsilon)=\mathcal{A}_1+\mathcal{A}_2\varepsilon+\mathcal{A}_3\varepsilon^2+....
\end{equation}
where the coefficients $\mathcal{A}_1$, $\mathcal{A}_2$ and $\mathcal{A}_3$ are defined as follows:
\begin{eqnarray}
\mathcal{A}_1&=&\left[\mathcal{T}_0+\frac{(\pi T)^2 (N-1)}{2}\left(\phi^2_1\cos2\delta_0-\sin2\delta_0\frac{\phi_2}{6}\right)\right], \\
\mathcal{A}_2&=&\alpha_1\sin2\delta_0,\\
\mathcal{A}_3&=& \left[
\cos2\delta_0\left(\alpha^2_1+\frac{(N-1)\phi^2_1}{2}\right)+\sin2\delta_0\alpha_2\right].
\end{eqnarray}
The transport integrals for $n=0, 1, 2$ are given by equations: 
\begin{equation}\label{j0j1new}
\mathcal{I}_0=\frac{N}{h}\left[\mathcal{A}_1+\frac{\mathcal{A}_3}{3}(\pi T)^2\right], \qquad \mathcal{I}_1=\frac{N\mathcal{A}_2}{3h}(\pi T)^2, \qquad \mathcal{I}_2=\frac{N}{h}\left[\frac{\mathcal{A}_1}{3}(\pi T)^2+\frac{7\mathcal{A}_3}{15}(\pi T)^4\right].
\end{equation}

The transport coefficients: the electrical conductance $G(T)$, thermopower $S(T)$ and thermal conductance $K_e(T)$ read as: 
\begin{equation}\label{nn}
G=e^2 {\cal I}_0, \quad S=-\frac{1}{eT}\frac{{\cal I}_1}{{\cal I}_0}, 
\quad K_e=\frac{1}{T}\left[{\cal I}_2-\frac{{\cal I}^2_1}{{\cal I}_0}\right].
\end{equation}
The electronic contribution to the thermoelectric figure of merit $ZT$ and the normalized power factor $PF$ are expressed in terms of thermoelectric properties defined in Eq.\eqref{nn} via
$ZT=S^2GT/K_e $ and $PF=S^2G/G_0$.

\end{widetext}

\end{document}